\documentclass[preprint,amssymb, numerical, aip, pof]{revtex4-1}
\usepackage{graphicx}
\usepackage{amsmath}
\usepackage{color}
\usepackage{graphics}
\usepackage[USenglish]{babel}
\usepackage{varioref}
\usepackage[T1]{fontenc}
\usepackage{ae}
\usepackage{fancyhdr}
\usepackage[latin1]{inputenc}
\usepackage{subfigure}
\usepackage{color}
\usepackage{appendix}
\usepackage{psfrag}

\newcommand{\pd}[2] {\frac{\partial #1}{\partial #2}}

\renewcommand{\vec}[1]{{\bf#1}}

\newcommand{\ddiv}{\vec \nabla^{\,*} \cdot}  
\newcommand{\dgrad}{\vec \nabla^{\,*}}  
\newcommand{\dlapl}{\nabla^{*2}}

\newcommand{\vu}{\vec u}

\usepackage{geometry}
\usepackage[colorlinks]{hyperref}

\begin{document}

\title{Drag reduction induced by superhydrophobic surfaces in turbulent pipe flow}
\author{Costantini R.}
\affiliation{Department of Mechanical and Aerospace Engineering, {\it Sapienza} 
University of Rome, via Eudossiana 18, 00184, Rome, Italy}
\author{Mollicone J.-P.}%
\affiliation{Department of Mechanical and Aerospace Engineering, {\it Sapienza} 
University of Rome, via Eudossiana 18, 00184, Rome, Italy}
\author{Battista F.}
\affiliation{Department of Mechanical and Aerospace Engineering, {\it Sapienza} 
University of Rome, via Eudossiana 18, 00184, Rome, Italy}
\email[]{francesco.battista@uniroma1.it}
\date{\today}%

\begin{abstract}
The drag reduction induced by superhydrophobic surfaces is investigated in turbulent pipe flow. 
Wetted superhydrophobic surfaces are shown to trap gas bubbles in their 
asperities. This stops the liquid from coming in direct contact with the wall in that location,
allowing the flow to slip over the air bubbles. We consider a well defined texture  
with streamwise grooves at the walls in which the gas is expected to be entrapped.
This configuration is modelled with alternating no-slip and shear-free boundary conditions at the wall. 
With respect to classical turbulent pipe flow, a substantial drag reduction is observed 
which strongly depends on the grooves' dimension and on the solid fraction,
i.e. the ratio between the solid wall surface and the total surface of the pipe's circumference.
The drag reduction is due to the mean slip velocity at the wall which increases  
the flow rate at a fixed pressure drop. The enforced boundary conditions also produce peculiar 
turbulent structures which on the contrary decrease the flow rate. 
The two concurrent effects provide an overall flow rate increase as demonstrated by means of the
mean axial momentum balance. This equation provides the balance between the mean pressure gradient,
the Reynolds stress, the mean flow rate and the 
mean slip velocity contributions.
\end{abstract}
\pacs{PACS}
\maketitle

\section{Introduction}  
\label{sec:intro}   

Many engineering devices are characterised by a solid wall in contact with a moving fluid.
The drag generated by the contact between fluid and walls affects the engineering system with 
significant energy and economic consequences \cite{frohn}. Drag reduction in turbulent flow can 
be achieved through various mechanisms, including the addition of polymers to the fluid \cite{white}, 
the addition of an air layer \cite{ceccio} or using riblets \cite{spalart}.

Recent advancements in nano- and micro- technologies have opened the possibility to create new surfaces 
with nano- or micro-scale roughness called superhydrophobic surfaces. Water droplets on a 
rugged surface typically exhibit one of the following two states \cite{lafuma}: (i) Wenzel State 
\cite{wenzel}, in which the water droplets are in full contact with the rugged surface; (ii) Cassie-Baxter 
State \cite{cassie}, in which water droplets are in contact with the peaks of the rugged surface 
and an ``air pocket'' is trapped between surface grooves. Recent studies showed the ability of these 
surfaces to induce substantial drag reduction when liquid flows over them both in laminar~\cite{ou} 
and in turbulent~\cite{roth} regimes. The moving fluid can ``slip'' in some areas at the wall where gas 
bubbles are entrapped in the surface asperities where normally, for an ordinary smooth surface, a zero 
slip velocity would be present \cite{pimponi}.  
In numerical computations, the superhydrophobic effect is typically modelled by alternating no-slip and 
free-shear boundary conditions. Depending on the placement of these boundary conditions, 
different types of patterns are obtained. Streamwise slip boundary conditions reproduce longitudinal 
grooves whilst spanwise slip boundary conditions reproduce transverse grooves. It follows that slip 
boundary conditions in both directions reproduce a combination of both patterns. 

The effect that these three possible superhydrophobic boundary conditions, i.e. 
(1) streamwise slip, (2) spanwise slip and (3) slip in both directions, have on the skin-friction drag in 
turbulent channel flow is shown by \citet{min} through a number of direct numerical simulations (DNS)  
performed at constant mass flow rate. The drag reduction increases proportionally with the slip length 
for case (1) whilst there is a decrease for case (2). For case (3), the reduction is less than 
case (1), due to the drag-increasing effect of the spanwise slip. \citet{min} state that the 
drag reduction is due to the streamwise slip-boundary condition which is a direct consequence of a smaller wall-shear stress.
The drag reduction/increase by 152 spanwise/streamwise slip length combinations 
is studied by \citet{busse} by imposing Navier-slip boundary conditions on turbulent channel flow.
The authors evaluate $\left( L_x, L_y \right)$-combinations that give no change in drag, 
then draw neutral curves that separate
drag-reducing and drag-increasing slip-length combinations. 

\citet{martell} use DNS to investigate a 
turbulent channel flow with superhydrophobic boundary conditions at the bottom wall 
for streamwise ridges or posts, expanding on \citet{min}'s work.
Another theoretical study by \citet{fukagata} is in good agreement with the results 
of \citet{min} and shows a clear correlation between drag reduction mechanism and Reynolds number. 
The code by \citet{min} is used by \citet{park} to investigate the internal flow through 
a superhydrophobic channel in both laminar and turbulent regimes.
The studies were performed by varying the spacing or gas fraction of microgrooves. 
Their results show that the drag reduction in turbulent flow depends on both the superhydrophobic patterns of the wall 
and the Reynolds number whilst in the laminar case a clear dependence on the former is observed.
\citet{rastegari2015}, through their exact analytical expression for the magnitude of drag reduction in channel flows, 
compare the behaviour of periodic superhydrophobic patterns (longitudinal microgrooves, transverse microgrooves and micro-posts) 
in laminar and turbulent regimes.
The authors separate the contribution of drag reduction arising from the effective slip on the wall 
from that due to the modification of the turbulence dynamics within the flow. 

Considering the wall structures developed in the turbulent regime,
the DNS of \citet{turk} is used to analyse secondary flow of Prandtl's second kind. 
This secondary motion consists of a pair of counter-rotating eddies 
that cover the entire channel height. 
The strength of the eddies increases with increasing spanwise dimension of grooves. 
This vortical motion transports fluid downwards over the free-slip region and upwards over the no-slip region, 
weakening the stability of the Cassie state. 
\citet{stroh} also study the evolution and the organisation of wall structures over
superhydrophobic surfaces for different solid fractions and wave lengths of shear-free/no-slip regions.  
\citet{im2017comparison} perform a comparison between a turbulent pipe and a turbulent
channel flow at a friction Reynolds number of 180 with a constant mass flow rate. The authors show that 
the drag reduction is higher in the pipe flow compared to the channel flow in analogous conditions.

Experimentally, \citet{tian} use time resolved particle image velocimetry (TRPIV) to show the geometry of 
hairpin vortices and compare those generated at superhydrophobic surfaces with the ones at hydrophilic surfaces. 
\citet{henoch} present surface fabrication techniques and analyse  
two superhydrophobic plates with different pattern configurations: 
a polyvinyl chloride (PVC) dummy plate and a nanograss plate, both in turbulent regimes. 
They measure higher drag reduction over 1.25 $\mu m$ spaced nanograss posts compared to the PVC plate. 
This result is confirmed over the entire range of speeds used for their tests. 
More recently, \citet{daniello} show a particle image velocimetry (PIV) study of  
a turbulent channel flow with two different superhydrophobic micro-ridge geometries in the streamwise direction 
at the bottom wall. The channel is tested over a range of mean Reynolds numbers (from $2000$ to $9500$)  
to investigate the effect of pattern changes on the velocity profiles, slip length and drag reduction. 
They show that the magnitude of the slip velocity increases proportionally with the Reynolds number and 
a maximum drag reduction of about $50 \%$ is observed for both cases. 
It is worthwhile to consider that the drag reduction in turbulent flow is only possible when the Cassie state occurs, 
thus making the stability of this state an essential factor.
Several studies~\cite{amabili,alj} address the Cassie-Baxter state stability in static conditions on a surface with defects.
The dependence of the meta-stability on thermodynamic conditions such as pressure and on the defect shape is addressed.
The experimental work of \citet{alj} shows that the stability of the air layer trapped between the grooves is 
critical for the effective drag reduction when using superhydrophobic surfaces, especially for high Reynolds 
number turbulent flows. When the coated flat plates analysed in their experiment are dynamically sheared by water, 
air bubbles are removed from the surface. The depletion of air is more intense with increasing Reynolds number. 
The superhydrophobic effect is completely lost for sufficiently large grooves when the Cassie-Baxter state 
is lost and the surface is fully wetted. 

The present work investigates the mechanism of drag reduction in a fully developed turbulent pipe flow with 
various superhydrophobic patterns at the highest Reynolds number (for this configuration) currently 
available in the literature. 
The analysis involves varying the geometry parameters of the wall pattern to study the effect they 
have on the overall drag reduction and how they modify turbulent structures.  We compare the mean 
flow rate for different cases, including a reference no-slip classical pipe flow. The phase average of 
radial velocity and axial vorticity reveals the existence of permanent structures in the near-wall region that 
strongly influence the slip velocity over the groove and wall shear stress over the wall. The overall 
momentum balance sheds light on the different contributions to the flow rate (hence the drag reduction).

Section 1 explains the numerical methodology on which the code is based and highlights the 
characterisation of the case study in terms of geometry, numerics and control parameters,
Section 2 shows the results of all the variables observed 
and Section 3 ends the paper with the final remarks. 

\section{Numerics}
\label{sec:numerics}

\begin{figure}[h!]
\centering
\includegraphics[width=0.9\textwidth]{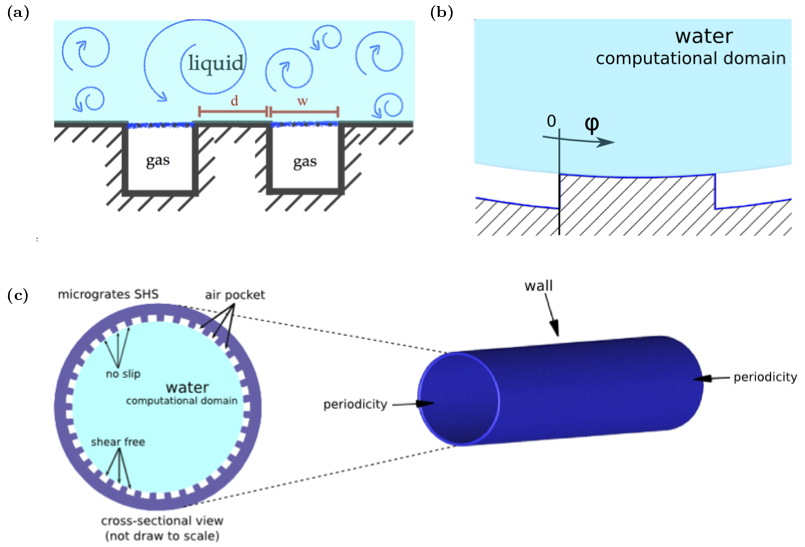}
\caption{\label{fig:geometry}  Description of the computational domain. 
Panel (a): sketch of the textured wall where the wetted surfaces have streamwise grooves in which gas bubbles are entrapped. 
Panel (b): detail of a wall post between two grooves. $\varphi$ is a local curvilinear coordinate. The cyan region
is the computational domain filled with liquid and the white regions represent the gas entrapped in the grooves.
Panel (c): cross-sectional view of the pipe. Cyan colour highlights the computational domain surrounded by the pipe's 
grooved wall which is enforced by alternating perfect-slip and no-slip alternated boundary conditions. 
The left part of the panel shows the actual geometry of the pipe with the boundary conditions set in the numerical simulation.
}
\end{figure}

The direct numerical simulations (DNS) of fully turbulent pipe flow driven by a constant pressure gradient 
are carried out. The boundary conditions and computational domain are shown in fig~\ref{fig:geometry}, where
$z$, $r$ and $\theta$ are the axial (stream-wise), radial (wall-normal) and azimuthal directions, respectively. 
The pipe walls are decorated with alternating no-slip/perfect-slip boundary conditions and periodic 
conditions are enforced in the stream-wise direction. A reference simulation with a fully no-slip wall condition is also
provided. The continuity and momentum equations, describing the evolution of a Newtonian fluid in the incompressible 
regime, read
\begin{align}
\label{eq:cont}
\ddiv{\vu^*} &= 0\, ,\\
\label{eq:mom}
\pd{\vu^*}{t^*} + \ddiv {\left(\vu^*\otimes\vu^*\right)} &= - \frac{\dgrad p^*}{\rho^*} +\nu^* \dlapl{\vu^*}\, ,
\end{align}
where $\vu^*$ is the fluid velocity and $\rho^*$, $p^*$, and $\nu^*$ are the density, hydrodynamic pressure, 
and kinematic viscosity respectively. The symbol ``$\otimes$'' denotes the tensor product and the asterisk 
denotes the dimensional quantities. Two sets of reference quantities are used to obtain the 
non-dimensional form of the data. The first set consists of the pipe radius $R^*$ and the bulk velocity 
$U_b^* = 2/R^{*2} \int_0^{R^*} r^* \langle u_z^*\rangle dr^*$ as the length and velocity reference scales, respectively 
(the angular brackets denote the ensemble average). The second set consists of the wall unit reference 
quantities, namely the viscous length $y_\tau^*=\nu^*/u_\tau^*$ and the friction velocity 
$u_\tau^*=\sqrt{\tau_w^*/\rho^*}$, where $\tau_w^* = \mu^* \left.d{\langle u_z^*\rangle}/d{y^*}\right|_{r^*=R^*}$ is the wall shear 
stress. Hereafter, the non-dimensional quantities in wall units are denoted with '$+$' superscripts whilst in external units no symbols are
used. The wall normal distance is $y^*=R^*-r^*$, $y=1-r$ in external units and $y^+ = (R^* - r^*)u_\tau^*/\nu^*$ in wall units.
The mass flow in the pipe is obtained through a mean pressure gradient in the axial direction, with pressure 
$p=\tilde{p}+(\Delta P/ L_z) z$, where $L_z= 2\pi$ is the overall length of the pipe, expressed in external units.
The pressure gradient is kept constant across all the simulations in order to obtain a specific friction Reynolds
number $Re_\tau = u_\tau^* R^* / \nu^*= 320$, which in the reference case corresponds to a bulk Reynolds number 
${\rm Re}=U_b^* D^* /\nu^* =10000$ based on the pipe diameter $D^* = 2\,R^*$. The various  
wall conditions are expected to produce mass flow variations.
This setup, also chosen by \citet{turk}, allows the investigation of drag reduction 
based on the difference in mass flow rate of the superhydrophobic surface cases with respect to
the no-slip reference pipe at the same $Re_\tau$. The simulations, under constant flow rate, can hold flow 
re-laminarisation effects if high drag-reduction values, possibly realised in superhydrophobic cases, are achieved. 
Therefore, the wall conditions in the superhydrophobic simulations are expected to produce a gain in mass flow rate variations 
with respect to the reference smooth one. 
The equation system~\eqref{eq:cont}-\eqref{eq:mom} in cylindrical coordinates is solved using a second-order scheme 
on a staggered grid with a local volume-flux formulation, see~\cite{battista2014turbulent,battista2015fractal,rocco2015curvature}. 
Both convective and diffusive terms are explicitly integrated in time using a third order Runge-Kutta 
low-storage method. The classical projection method is used to enforce the continuity 
equation~\eqref{eq:cont} constraint.  The computational domain is shown in figure~\ref{fig:geometry}(c) whilst the
collocation points employed for simulations are reported in the second column of table~\ref{tab:sim}. The grid 
spacing is constant in the azimuthal and axial directions whilst it is reduced in the near wall region in the radial direction 
to satisfy the resolution requirements, $ \Delta r \simeq y^+ = 0.003$. 
MPI (Message Passing Interface) directives are employed for parallel computing and the
two-dimensional pencil decomposition is implemented through the 2decomp\&fft libraries~\cite{2decomp}.

\begin{figure}[h!]
\centering
\includegraphics[width=.9\textwidth]{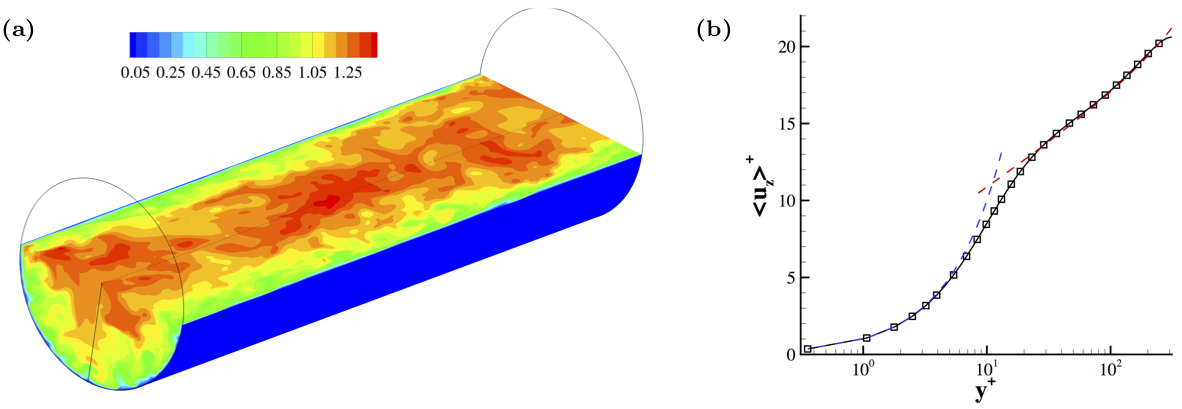}
\caption{\label{fig:instant} 
Panel (a): instantaneous axial velocity of the turbulent periodic pipe flow 
for the reference simulation with no-slip boundary conditions. 
Panel (b): plot of mean streamwise (z-direction) velocity normalised with 
friction velocity $\langle u_z\rangle^+=\langle u_z\rangle/u_\tau$ against $y^+$ for the same simulation in panel (a). 
The dashed blue line is the theoretical prediction $\langle u_z\rangle^+=y^+$ in the 
viscous sub-layer. 
The dashed red line is the fit 
$\langle u_z\rangle^+ = {1}/{k} \log{(y^+)} + {g \, y^+}/{Re_{\tau}}  + A$
in the log-layer region, with $k=0.392$ and $A=4.5$ ($g$ is a constant equal to $2$ for the pipe flow). 
The effect of finite Reynolds number is accounted for as formulated in~\cite{Luchini2017}.}
\end{figure}

As anticipated, the wall is set with alternating no-slip/shear-free boundary conditions to mimic the presence of 
streamwise aligned ridges of width $d$, alternated with grooves of width $w$ where the gas phase is entrapped. 
A liquid-gas interface is pinned at the edge of the grooves, representing a Cassie-Baxter stable state, as shown 
in panel (a) of figure~\ref{fig:geometry}. The no-slip and the no-penetration boundary conditions are enforced 
on the ridge, while the shear-free and the no-penetration boundary conditions on the liquid-gas interfaces, 
\begin{align}
u_r=0, \quad& u_\theta=0, \quad u_z=0 \quad \quad \text{on ridges}\\
u_r=0, \quad& \pd{u_\theta}{r}=0,	\quad \pd{u_z}{r}=0 \quad \quad \text{on liquid-gas interface}\, .
\end{align}
In these conditions, the interface is a fixed boundary on which a perfect slip condition is enforced. The computational domain
therefore consists of the cyan part in figure~\ref{fig:geometry}(c).
A reference simulation is performed with no-slip boundary conditions and the same friction Reynolds number of
the other simulations, $Re_\tau= 320$, with a corresponding nominal Reynolds number $Re = 10000$. 
Figure~\ref{fig:instant}(b) reports the semi-logarithmic plot of the mean velocity profile in wall units (symbols) 
with the theoretical trends~\cite{pope2001turbulent} in both the viscous sub-layer and the log-layer superimposed as dashed lines. 
Note that the effect of the pressure gradient is accounted for in the plots in the form of a higher-order perturbation~\cite{Luchini2017}. 
The agreement between the simulation data and theoretical expectation confirms that the tuned resolution is suitable to capture the 
near wall turbulent structures. Hereafter, the reference simulation will be used to compare the effects of the free-slip boundary 
conditions on the turbulent structures and features.

Various simulations have been performed to see the effect of superhydrophobic wall patterning 
on the turbulent structures near the boundary and on the overall drag reduction/increase.
The features of the simulations are summarised in table~\ref{tab:sim}. 
The groove width, $w$, is changed whilst the solid fraction, i.e. the solid surface to the overall surface ratio, is kept constant at 
$\Phi_S=0.5$. On the other hand, for cases C and D, additional simulations are performed by keeping the periodicity constant, 
$L =d+w$, and changing the solid fraction.

\begin{table}
\centering
\setlength{\tabcolsep}{10pt}
\begin{tabular}{cccccccccc}
SIM    & Grid size                   &$L$    &$L^+$  & $\Phi_S$ & $\langle u_s\rangle$ & $U_b$ & $C_f$ & $l^+_s$\\ \hline
REF   & $384 \times 129 \times 256$ &$-$    & $-$   & 1.00     & $-$  & 1.000   & 0.0077 & $-$ \\
A50   & $768 \times 129 \times 256$ & 0.066 & 21.0  & 0.50     &0.210 & 1.136   & 0.0060 & $3.395$\\
B50   & $768 \times 129 \times 256$ & 0.098 & 31.4  & 0.50     &0.249 & 1.146   & 0.0059 & $4.014$ \\
C50   & $384 \times 129 \times 256$ & 0.130 & 41.9  & 0.50     &0.282 & 1.151   & 0.0058 & $4.544$ \\
D50   & $384 \times 129 \times 256$ & 0.262 & 83.8  & 0.50     &0.404 & 1.223   & 0.0051 & $6.479$ \\
E50   & $384 \times 129 \times 256$ & 0.524 & 167.6 & 0.50     &0.538 & 1.357   & 0.0042 & $8.608$ \\ 
C75   & $768 \times 129 \times 256$ & 0.130 & 41.9  & 0.75     &0.088 & 1.071   & 0.0067 & $1.410$ \\ 
C25   & $768 \times 129 \times 256$ & 0.130 & 41.9  & 0.25     &0.694 & 1.453   & 0.0036 & $11.233$ \\ 
D75   & $384 \times 129 \times 256$ & 0.262 & 83.8  & 0.75     &0.126 & 1.060   & 0.0068 & $2.026$ \\ 
D25   & $384 \times 129 \times 256$ & 0.262 & 83.8  & 0.25     &0.935 & 1.617   & 0.0029 & $15.032$ \\ 
\hline
\end{tabular}
\caption{\label{tab:sim} Simulation parameters. From left to
right: simulation codename; grid size in azimuthal, radial and 
axial directions respectively; periodicity length in external units, $L$, and in 
wall units, $L^+=L/y_{\tau}$; solid fraction $\Phi_S$; mean slip velocity at the wall $\langle u_s\rangle$; bulk velocity $U_b$;  
friction coefficient $C_f$ and effective slip length in wall unit, $l_s^+$. 
}
\end{table}

\subsection{Statistical tools}
\label{sec:phase}

Figure~\ref{fig:instant} shows the instantaneous axial velocity $u_z$ of the reference simulation REF which has only no-slip 
boundary conditions at the wall. For the statistical analysis, 200 uncorrelated instantaneous fields are collected, taken at 
every $t^*=2.5 \, t_0^*$, where $t_0^*=R^*/U_b^*$ is the reference time. 
The flow is statistically stationary and homogeneous in the axial and azimuthal direction. The homogeneity in the 
azimuthal direction is evidently satisfied by the physical geometry of the pipe whilst in the streamwise direction  
the periodicity of the boundary conditions suggest the existence
of statistical homogeneity. Suitable phase averaging allows a triple decomposition~\cite{reynolds1972mechanics, stroh}.

A local curvilinear coordinate $\varphi$ is defined for the periodic pattern in the 
azimuthal direction~\cite{stroh}, see figure~\ref{fig:geometry}(b). The phase average of an arbitrary quantity $q$, indicated by the angular brackets, can be calculated by the equation 
\begin{equation}
\langle q| \varphi \rangle (\varphi,r) = \frac{1} {{N L_z} T} \sum_{n=1}^N \int_t \int_z q\left( \frac{\varphi}{2\pi}+n,r,z,t\right)dzdt\, ,
\label{eq:mean_ph}
\end{equation}
where $T$ is the time during which the instantaneous fields are 
collected and $N$ is the number of the wall-groove couples in the azimuthal direction. The average 
over the grooves, $\langle u_z|\varphi_g\rangle$, or over the stripes, $\langle u_z|\varphi_w\rangle$, 
is defined as 
\begin{align}
\langle q| \varphi_{g/w} \rangle (\varphi_{g/w},r) = \frac{1}{\Theta_{g/w}}\int_\varphi \langle q| \varphi_{g/w} \rangle (\varphi,r) d\varphi \, ,
\label{eq:mean_ph1}
\end{align}
where $\Theta_{g/w}$ is the grooves/stripes width. The classical spatial mean can be obtained by 
integrating over the phase coordinate $\varphi$,
\begin{equation}
\langle{q}\rangle (r) = \frac{1}{2\pi}\int_0^{2\pi} \langle q | \varphi \rangle (\varphi,r) d\varphi\, .
\label{eq:mean}
\end{equation}
The corresponding decomposition is 
\begin{equation}
\tilde{q} (\varphi,r) = \langle q|\varphi\rangle (\varphi,r) - \langle {q}\rangle(r)\, ,
\label{eq:dec1}
\end{equation}
and therefore any flow variable can be decomposed as 
\begin{align}
\notag q(\theta,r,z,t) &=  \langle q|\varphi\rangle (\varphi,r) + q'' (\theta,r,z,t) =\\
\notag &= \langle{q}\rangle(r) + \tilde{q} (\varphi,r) + q'' (\theta,r,z,t) =\\
&= \langle{q}\rangle(r) + q' (\theta,r,z,t)    \, ,
\label{eq:dec1}
\end{align}
where the double prime denotes the fluctuation with respect to the phase average whilst the single prime denotes
the classical fluctuation with respect to the overall spatial mean which is equivalent to the fluctuation in the conventional 
Reynolds decomposition. 

\section{Results}
\label{sec:results}

\begin{figure}[h!]
\centering
\includegraphics[width=1\textwidth]{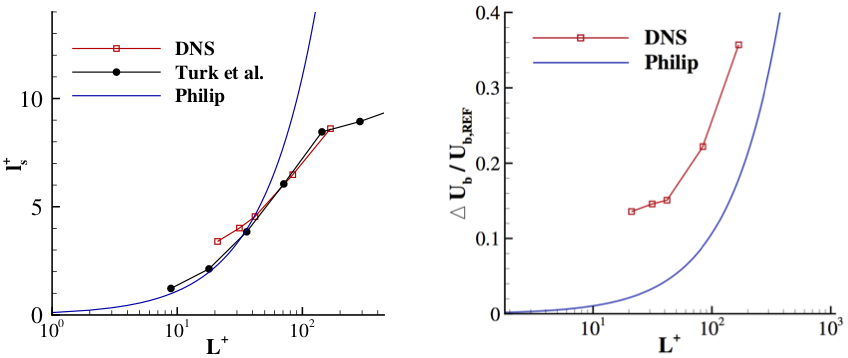}
\caption{\label{fig:ls_ub} 
Left panel: slip length $l_s^+$ comparison between current DNS data, \citet{turk} data 
and \citet{philip1972integral}'s laminar solution as a function of the periodicity length $L^+$. 
Right panel: comparison of flow rate increase $\Delta U_b / U_{b,REF}$ between current DNS data
and \citet{philip1972integral}'s laminar solution as a function of the periodicity length $L^+$. 
From left to right, the red open squares represent simulations A50, B50, C50, D50 and E50. } 
\end{figure}
The simulations are performed at a fixed pressure gradient $\Delta P/L_z$ and therefore 
the boundary conditions directly affect the velocity profiles. 
The flow rate, represented through the bulk velocity since the pipe cross section is constant in all simulations, and the slip length, 
summarised in table~\ref{tab:sim}, can be compared with the theoretical values calculated in laminar regime
for which there exist analytical equations.
Philip's solution \cite{philip1972flows, philip1972integral}, also applied to pressure-driven Stokes flow by \citet{lauga},
shows the velocity increment in a circular laminar tube with 
$N$ longitudinal no-shear strips with respect to the Poiseuille solution of a canonical pipe with no-slip boundary condition.
The slip length $\left. l_s \right|_{lam}$ is deduced from this increment, which is a consequence of the slip velocity.
The increase in flow rate 
$\left. \left( \frac{\Delta U_b}{U_{b,REF}} \right) \right|_{lam}$ in this laminar regime 
provides a link between the geometrical structure of the surface pattern, $L$, and solid fraction, $\Phi_S$:
\begin{align}
\label{eq:ls_lam}
\left. l_s \right|_{lam} = \frac{L}{\pi} \log \{ \cos^{-1} \left[ \frac{\pi}{2} \left( 1-\Phi_S \right) \right] \}
\end{align}
\begin{align}
\label{eq:ub_lam}
\left. \left( \frac{\Delta U_b}{U_{b,REF}} \right) \right|_{lam} = \frac{-\frac{L}{\pi} \log \left[ \cos^{-1} \left[ \frac{\pi}{2} \left( 1-\Phi_S \right) \right] \right] }{Re_{\tau} / 3}
\end{align}
The left panel of figure \ref{fig:ls_ub} shows the slip lengths, reported in table \ref{tab:sim}, compared to
equation~\ref{eq:ls_lam} evaluated for the simulations with constant solid fraction $\Phi_S=0.5$ and 
the slip length obtained by \citet{turk} in a channel flow configuration with similar boundary conditions at 
$\textrm{Re}=180$.
DNS cases exhibit very similar behaviour with only a slight deviation for simulations with periodicity length under the value of $L^+=41.9$. 
The simulations by \citet{turk} coincide with the laminar solution below $L^+=35$, indicating that $\left. l_s^+ \right|_{turb}$ of the channel
is mainly a function of the geometrical properties of the streamwise grooves.
On the other hand, current DNS data always differs from the laminar solutions except for case C50 in which there is 
a reversal of the trend: under $L^+=41.9$, $\left. l_s^+ \right|_{lam} < \left. l_s^+ \right|_{turb}$, whilst 
simulations D50 and E50 deviate from the theoretical laminar solution with $\left. l_s^+ \right|_{lam} > \left. l_s^+ \right|_{turb}$. 
This feature indicates that the slip length of the superhydrophobic pipe always depends on the flow regime. 
The slip length is higher for A50 and B50 and lower for D50 and E50, compared with the laminar curve, indicating that
for the former two cases the turbulent regime intensifies the positive effects of the superhydrophobic surface in laminar conditions, opposed to 
the latter two cases which show an opposite trend. 
The right panel of figure \ref{fig:ls_ub} shows a comparison between the bulk velocity increase, and therefore the flow 
rate increase, for the laminar equation \ref{eq:ub_lam} 
and the present DNSs for constant solid fraction $\Phi_S=0.50$.
As shown in table \ref{tab:sim}, the bulk mean velocity of the superhydrophobic cases always increases 
with respect to the no-slip reference. This gain is also larger than the theoretical laminar trend 
studied by Philip.
Furthermore, the decrease in slip length for D50 and E50 discussed for the left panel of figure \ref{fig:ls_ub} does 
not affect the global gain in flow rate. 

\begin{figure}[h!]
\centering
\includegraphics[width=0.9\textwidth]{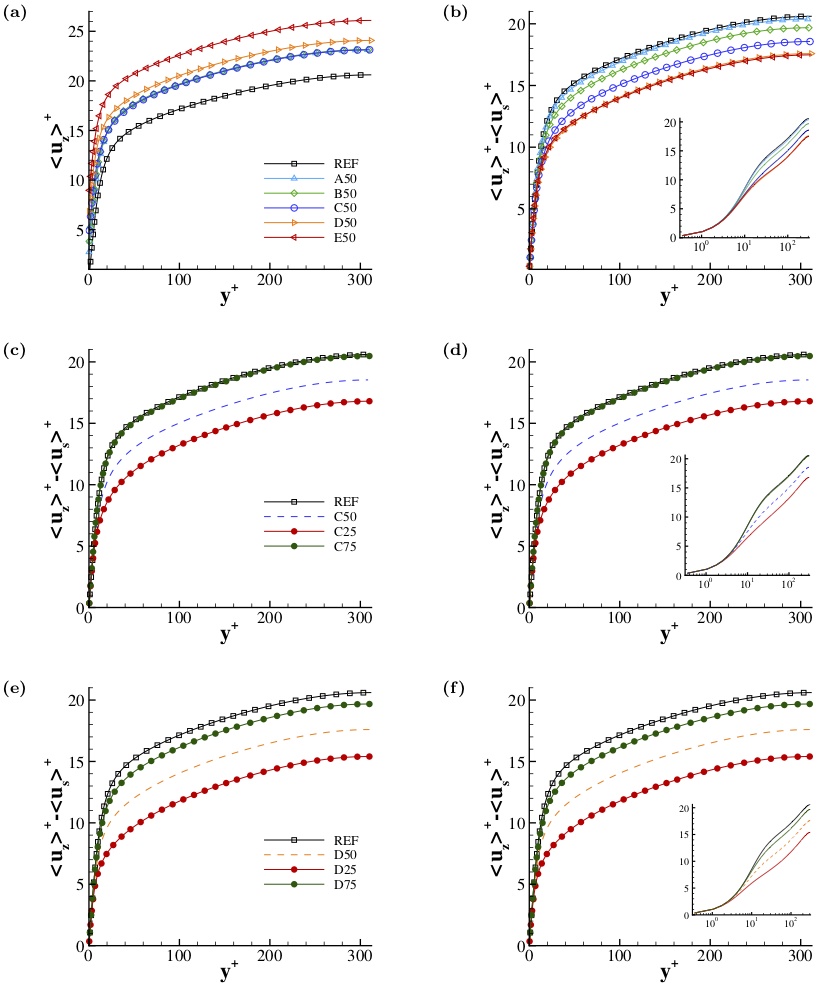}
\caption{\label{fig:uz} Normalised mean axial velocity $\langle u_z\rangle ^+$ against  
radial distance from the wall normalised with the wall viscous length scale $y^+$. Left panels: 
normalised mean axial velocity $\langle u_z \rangle /u_\tau$ (where $u_\tau=\sqrt{\tau_w/\rho}$ is the friction
velocity). Right panels: normalised mean axial velocity purged of the mean slip velocity,
$(\langle u_z \rangle- \langle u_s \rangle)/u_\tau$. The inset in panels (b), (c) and (d) is the semi-log representation of the same plot.
Note that for a clear plot the symbols only represent a subset of the total computational grid points
and a proper grid point clustering is employed at the wall to resolve well the turbulent structures.} 
\end{figure}

\subsection{Mean profiles} 

Panels (a), (c) and (e) in figure~\ref{fig:uz} show the mean axial velocity profiles 
$\langle u_z \rangle^+ (r)$ against the wall distance $y^+$ at different groove dimensions 
and fixed solid fraction $\Phi$ in panel (a) and at varying solid fraction and fixed periodicity $L$ in panels (c) 
and (e). The superhydrophobic boundary conditions produce higher axial velocity with respect to 
the reference no-slip case REF. This means there is a higher flow rate which corresponds to drag reduction, 
see panel (a). A measure of this drag reduction is given by the friction 
coefficient $C_f=2 \tau_w/\rho U_b^2$ which is reported in table~\ref{tab:sim}. 
The drag reduction ranges from 12\% for case A50 up to 62\% for case D25, compared to case REF. 
The percentages are overall high but this is probably due to the assumption that the Cassie-Baxter state 
is stable in our configuration and the liquid-gas interface is fixed. We expect that without these assumptions,  
these percentages would decrease.
In panel (a), the velocity increment is not monotonic with the dimensions of the grooves. 
Although the mean slip velocity $\langle u_s\rangle =\left. \langle u_z\rangle\right|_{r=1}$ varies depending
on the groove dimension, most of the superhydrophobic cases present a similar velocity profile. 
Simulation E50 shows significantly higher velocity and therefore a greater drag reduction.
Panels (c) and (e) show that the variation of the solid fraction has a strong effect on the flow rate 
which increases with increasing groove fraction, see cases C25 and D25. 
This behaviour is attributed to the higher slip velocity at the boundary due to the larger  
fraction of the liquid/gas interface. 

The present unconventional boundary conditions induce two phenomena:
the velocity increases due to the slip of the fluid at the liquid-gas interface 
at the grooves and the velocity fluctuations increase due to the alternating perfect-slip and no-slip regions.

Figure~\ref{fig:uz}(b) shows the radial profiles of the mean axial velocity minus the mean
slip velocity contribution, $\langle u_z \rangle^+ - \langle u_s \rangle^+$.  
The opposite behaviour is now observed with respect to the profiles in panel (a).
The values of $\langle u_z \rangle^+ - \langle u_s \rangle^+$ for the superhydrophobic cases are 
lower than the reference case and decrease proportionally with the increase in groove width. 
The modification of the turbulent fluctuations due to the boundary
conditions therefore produces an increase in drag with respect to the no-slip case. 
This effect is nonetheless not strong enough to overwhelm the velocity increase due to slip at the interface. 
The overall effect, as plotted in panel (a), therefore remains drag reducing. 
The inset of panel (b) reports the same velocity profiles in semi-log plot.
As expected, the profiles collapse in the viscous sublayer.
On the other contrary, in the log-layer region, they depart from the canonical pipe flow 
plots~\cite{pope2001turbulent} shown in the inset, indicating an increase in drag. 
Panels (d) and (f) confirm the previous observations, i.e. the larger the groove width, the higher the 
velocity fluctuations, therefore resulting in turbulence modification and consequently drag increase. 
\begin{figure}[h]
\centering
\includegraphics[width=1\textwidth]{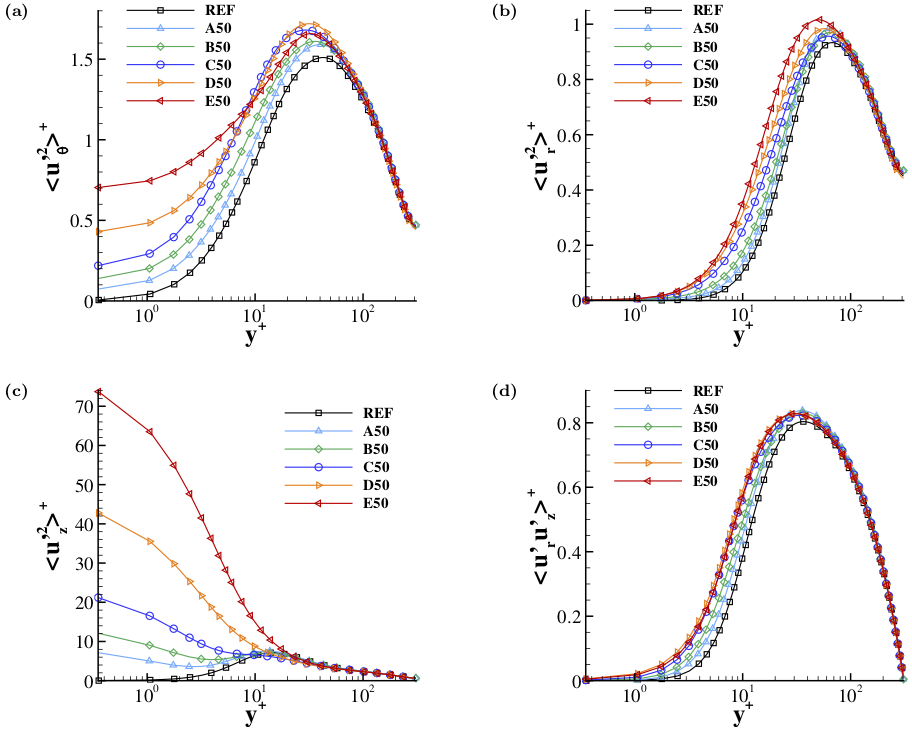}
\caption{\label{fig:Re} Reynolds stress tensor components 
against radial distance from the wall (in semi-logarithmic scale).
(a) Azimuthal component $\langle u^{'2}_{\theta} \rangle^{+} $
(b) Radial component $\langle u^{'2}_{r} \rangle^{+} $  
(c) Axial (streamwise) component $\langle u^{'2}_{z} \rangle^{+} $ 
(d) Reynolds stress $\langle u'_{r} u'_{z} \rangle^{+} $.
All the profiles and radial distance are expressed in wall units.
Note that for a clear plot the symbols only represent a subset of the total computational grid points
and a proper grid point clustering is employed at the wall to well resolve the turbulent structures.
}
\end{figure}

Figure~\ref{fig:Re} shows the radial profiles of the Reynolds stress tensor components normalised
with the friction velocity square. The azimuthal and axial components of the Reynolds stresses are showed in 
panel (a) and (c), respectively. $\langle u_{\theta}'^{2} \rangle^+$ and $\langle u_{z}'^{2} \rangle^+$ profiles show 
higher velocity fluctuations with respect to REF at the boundary (which is affected by the slip of the fluid at the 
liquid/gas interface). These velocity fluctuations increase proportionally with the width of the grooves. In the azimuthal 
direction, the fluctuation peak in the buffer layer at $y^+\sim 30$ increases and slightly shifts towards the wall. 
In the bulk region, the velocity fluctuations are unaffected by boundary.
The radial component of the Reynolds stresses, $\langle u_{r}'^{2} \rangle^+$, is zero at the boundary due to the
impermeability condition both at the wall and at the liquid/gas interface. In the buffer layer, the
radial velocity fluctuations progressively increase with the increase in the groove width. 
The shear component of the Reynolds stresses, $\tau_R = \langle u'_{r} u'_{z} \rangle^+$, is zero at the 
wall and shows higher values in the buffer layer for the superhydrophobic cases with respect to REF. 
The increase in $\tau_R = \langle u'_{r} u'_{z} \rangle^+$ is not monotonic with respect to the groove width 
and the maximum fluctuations are reached for the C50 case.

\begin{figure}[h]
\centering
\includegraphics[width=1\textwidth]{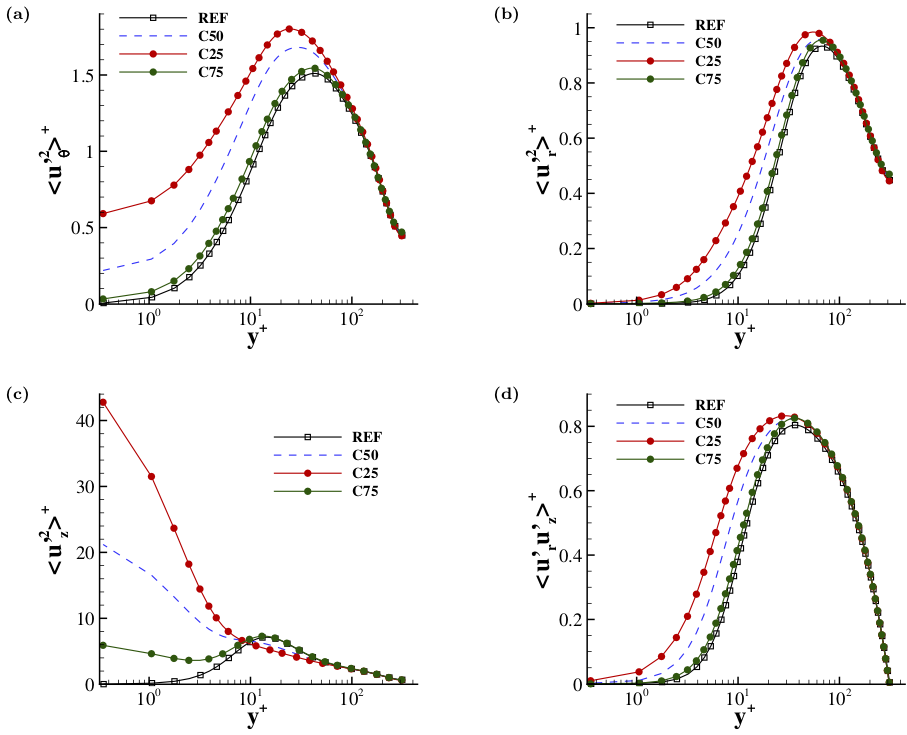}
\caption{\label{fig:Re_6-2} Reynolds stress tensor components  
against radial distance from the wall (in semi-logarithmic scale).
(a) Azimuthal component $\langle u^{'2}_{\theta} \rangle^{+} $
(b) Radial component $\langle u^{'2}_{r} \rangle^{+} $  
(c) Streamwise component $\langle u^{'2}_{z} \rangle^{+} $ 
(d) Reynolds stress $\langle u'_{r} u'_{z} \rangle^{+} $.
Note that for a clear plot the symbols only represent a subset of the total computational grid points
and a proper grid point clustering is employed at the wall to well resolve the turbulent structures.
}
\end{figure}
\begin{figure}[h]
\centering
\includegraphics[width=1\textwidth]{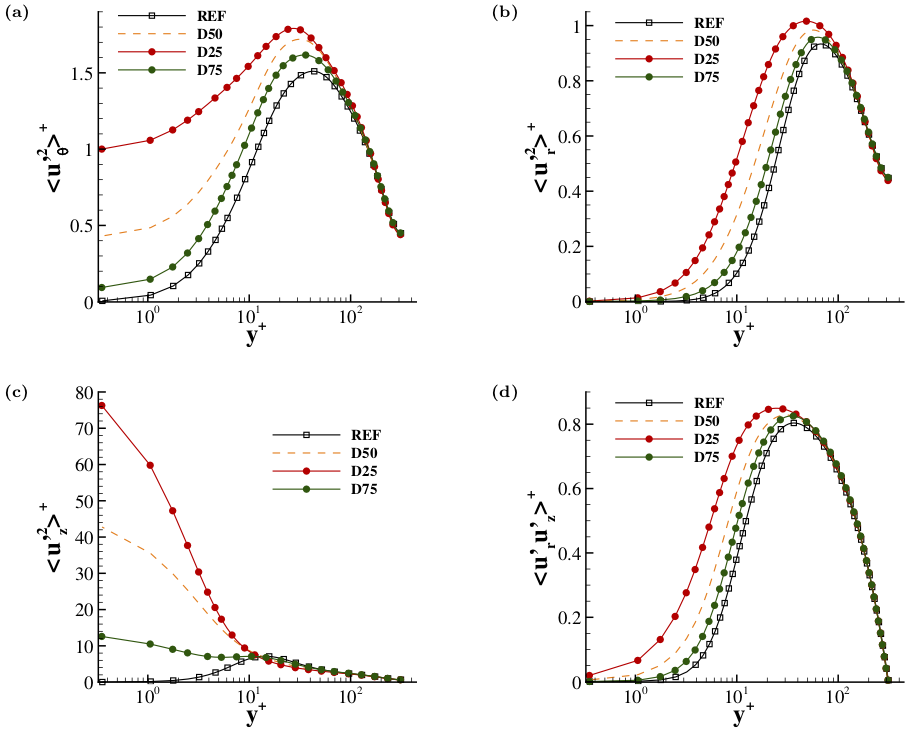}
\caption{\label{fig:Re_12-4} Reynolds stress tensor components  
against radial distance from the wall (in semi-logarithmic scale). 
(a) Azimuthal component $\langle u^{'2}_{\theta} \rangle^{+} $
(b) Radial component $\langle u^{'2}_{r} \rangle^{+} $  
(c) Streamwise component $\langle u^{'2}_{z} \rangle^{+} $ 
(d) Reynolds stress $\langle u'_{r} u'_{z} \rangle^{+} $.
Note that for a clear plot the symbols only represent a subset of the total computational grid points
and a proper grid point clustering is employed at the wall to well resolve the turbulent structures.
}
\end{figure}
\begin{figure}[h!]
\centering
\includegraphics[width=1\textwidth]{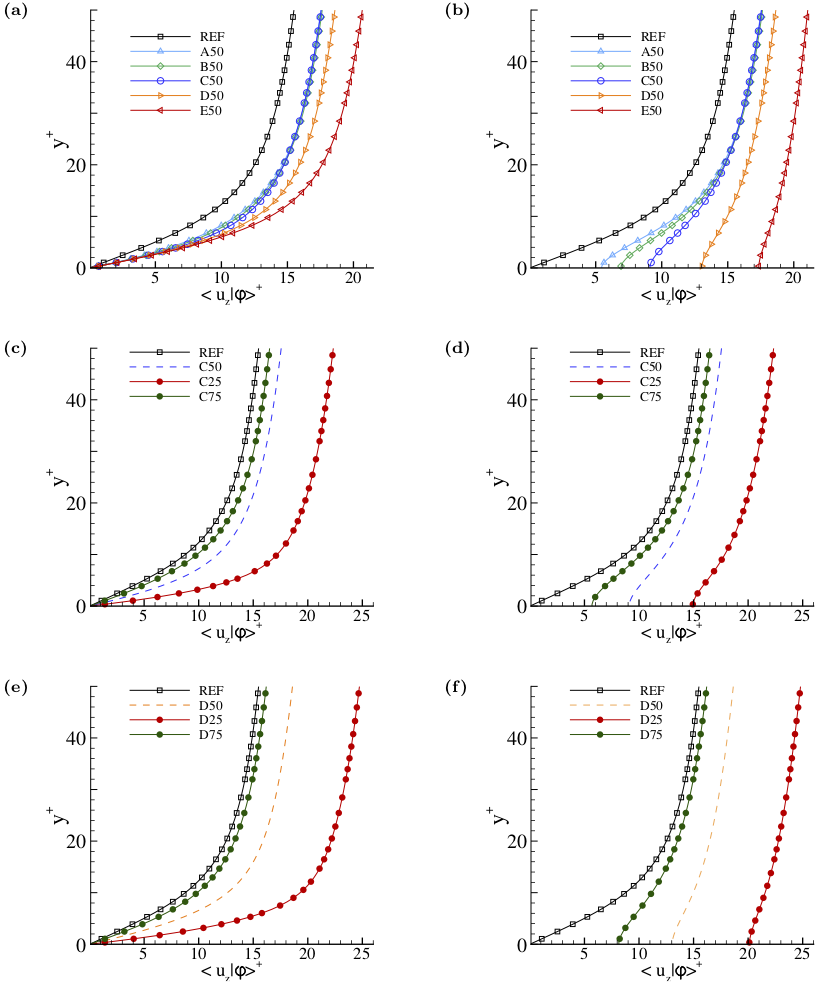}
\caption{\label{fig:uz_ph} Mean conditioned axial velocity 
against radial distance from the wall (both expressed in wall units).
Left and right panels report the profiles in correspondence of the no-slip and 
perfect slip boundary condition, respectively.
Note that again the symbols only represent a subset of the total computational grid points used 
and a proper grid point clustering is employed at the wall to well resolve the turbulent structures.
}
\end{figure}
Both figures~\ref{fig:Re_6-2} and~\ref{fig:Re_12-4} show the Reynolds stress components (normalised
with the square of the friction velocity $u_\tau$) against the radial distance from the wall as a function 
of the solid fraction. The qualitative behaviour is the same as the other 
superhydrophobic cases. Under a quantitative point of view, the decrease in solid fraction 
produces higher velocity fluctuations. The cases with smallest solid fraction 
(C75 and D75) have negligible differences with the reference case except for the axial velocity 
fluctuations $\langle u_z'^2\rangle^+$ which present a sensible departure close to the walls.
Figure~\ref{fig:uz_ph} shows the radial profile of the streamwise velocity phase average, 
see~\eqref{eq:mean_ph1}, coinciding with the solid wall, $\langle u_z |\varphi_w\rangle$ in  
panels (a), (c) and (e), and with the interface, $\langle u_z |\varphi_g\rangle$ in  
panels (b), (d) and (f), therefore dividing the no-slip region from the shear-free region. 
At the wall, the mean streamwise velocity is zero for all the cases in panel (a) since there is no interface. 
At the interface, panel (b), only the reference case has zero velocity and is drawn just as a reference plot. 
At the solid wall, the wall shear stress of the reference simulation
is lower than the superhydrophobic cases. This behaviour can be deduced from the mean streamwise momentum 
balance, $S_w \tau_w= \Delta p\, A$, where $S_w=2\pi R \Phi_S L_z$ is the solid wall surface and $A=\pi R^2$ is the 
cross-section area. 
{
For the superhydrophobic cases with $\Phi_S=0.5$, the solid wall surface is half that of the reference case 
and therefore the wall shear stress, $\tau_w$ has to double. 
The same $\tau_w$ is maintained since the total pipe boundary is constant 
for all these configurations. 
For the same reasons, the shear stress increases with the decrease of the solid fraction $\Phi_S$,
see panels (c) and (d). 
At the interface, the slip velocity increases proportionally with the groove width, 
panel (b), and decreases proportionally to the solid fraction, panels (d) and (f).
This behaviour increases the velocity in the bulk region even though it is not proportional 
to the groove dimensions (especially for small groove widths). }
\begin{figure}[h!]
\centering
\includegraphics[width=1\textwidth]{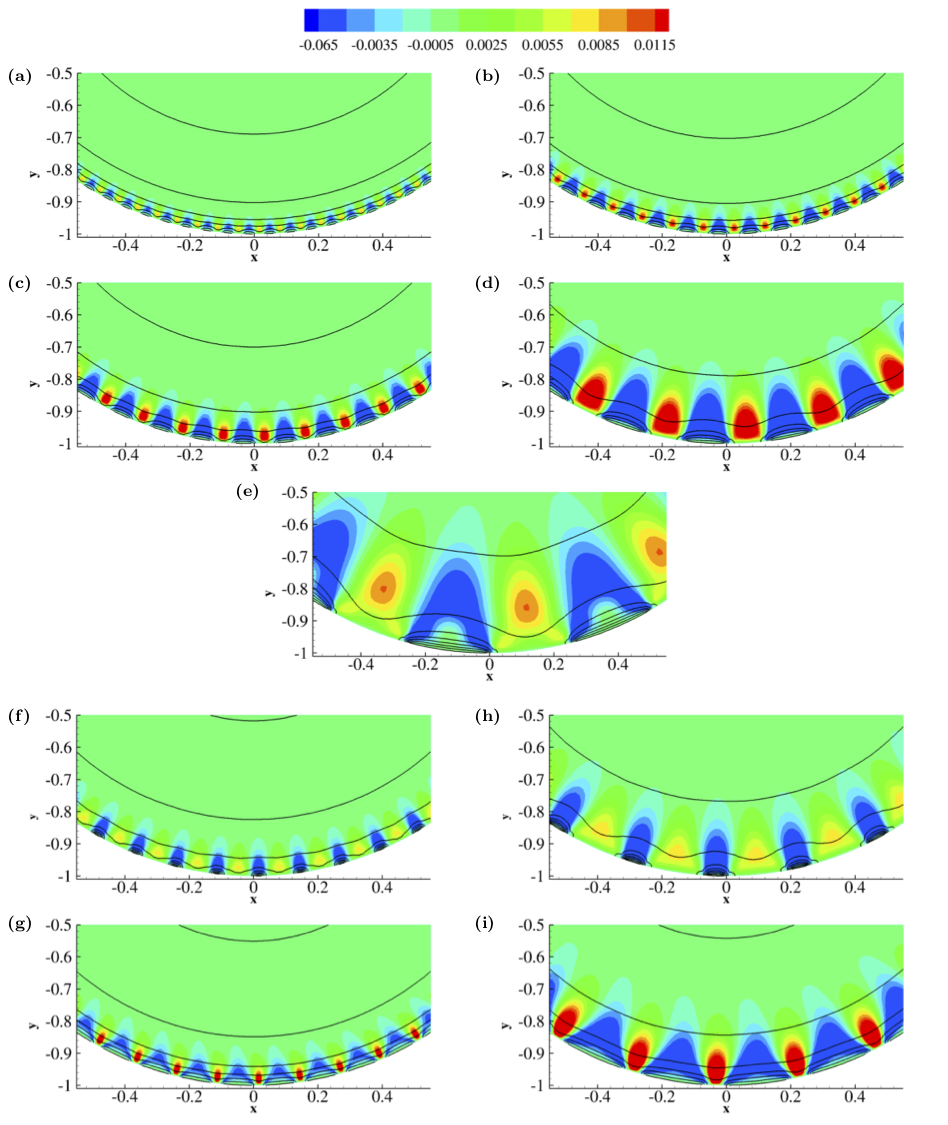}
\caption{\label{fig:ur_phase} Phase average of the radial velocity normalised with the mean slip velocity,
$\langle u_r|\varphi\rangle/\langle u_s\rangle$ as contour plots and 
phase average of the axial velocity  $\langle u_z|\varphi\rangle$ as solid lines for all the cases. 
Panels from (a) to (i) represent the cases in the order listed in table~\ref{tab:sim} 
(REF simulation is omitted).
}
\end{figure}

\subsection{Phase average statistics}

Figure~\ref{fig:ur_phase} shows the phase average radial velocity normalised with the corresponding 
mean slip velocity, $\langle u_r|\varphi\rangle/\langle u_s\rangle(\varphi,r)$ as coloured contour plots 
and the phase average axial (streamwise) velocity, $\langle u_z|\varphi\rangle(\varphi,r)$, as solid 
black lines. The superhydrophobic cases are shown, from panel (a) to panel (i), omitting 
the reference simulation whose mean radial velocity is uniformly zero. 
The black solid lines show significant changes in axial velocity at the boundary since the velocity has to necessarily 
be zero over the solid wall (no-slip boundary conditions). 
Over this solid wall, the mean radial velocity is negative and the fluid moves towards the pipe axis. 
On the other hand, the velocity is positive over the liquid/gas interface and the fluid 
moves towards the solid boundary. The same qualitative behaviour occurs for all the superhydrophobic 
cases. The data is normalised with the slip velocity to maintain the velocity range constant
for all the plots. 
On the other hand, consistently with the mean slip velocity in table~\ref{tab:sim}, 
the radial velocity increases with the dimension of the grooves up to D50, panel (d). A more 
complex behaviour occurs in case E50, panel (e). The groove dimension strongly influences the 
extension of the region in which flow modification occurs. Since the qualitative behaviour is the same for 
all the cases, we shall focus on case D50 to better characterise the turbulence modification induced by the 
alternating no-slip/free-shear boundary conditions.

\begin{figure}[h]
\centering
\includegraphics[width=\textwidth]{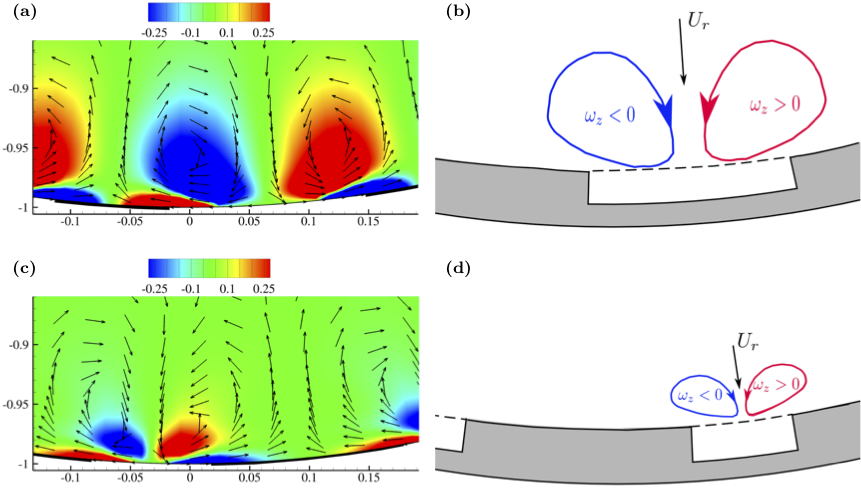}
\caption{\label{fig:phase_struct} Phase average of vorticity $\langle\zeta_z|\varphi\rangle$ represented 
by colours and the phase average of in-plane velocity components as vectors for cases D50 and D75 in 
panels (a) and (c) respectively. Black solid lines represent the walls. 
Panels (b) and (d) show sketches of the streamwise component of vorticity, with a 
graphical representation of the near-wall vortical structures induced by the alternating 
no-slip/free-shear boundary conditions for the corresponding cases in panels (a) and (c).}
\end{figure}

\begin{figure}[h]
\centering
\includegraphics[width=.49\textwidth]{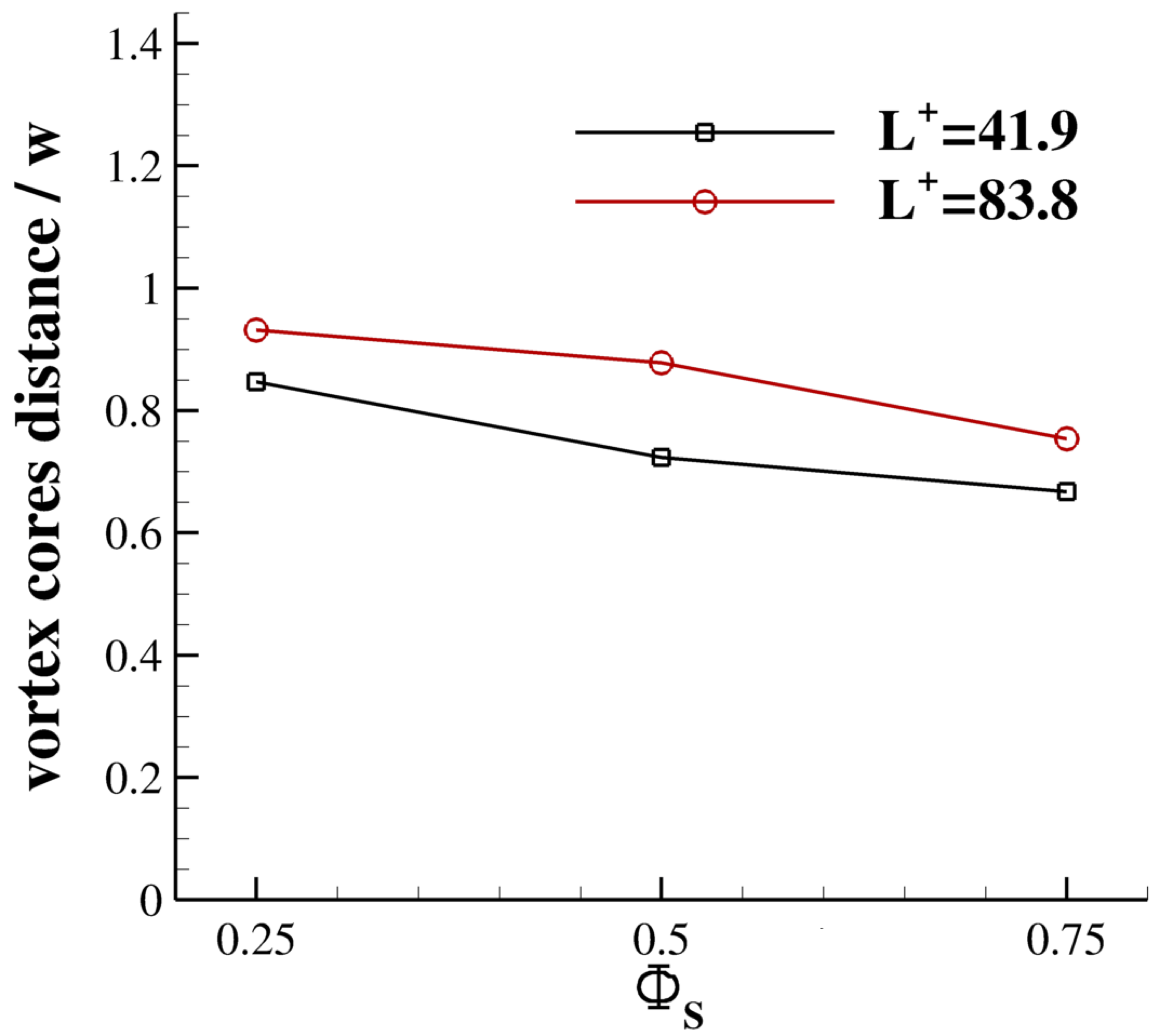}
\caption{\label{fig:core_dist} Distance between two mean vortices, related to figure~\ref{fig:phase_struct}, as a function of the solid 
fraction normalised with groove width. From left to right, simulations C25, C50 and C75 are in open black squares whilst simulations 
D25, D50 and D75 are in open red circles.
}
\end{figure}

Figure~\ref{fig:phase_struct} illustrates the mean structures near the wall for case D50. 
Panels (a) and (c) show the phase average of the streamwise component of 
vorticity $\langle\zeta_z|\varphi\rangle$ (coloured contour), which is the only non-zero mean vorticity component, 
and the phase average of the in-plane velocity components (black vectors) for cases D50 and D75 respectively. 
Panels (b) and (d) are sketches that provide a graphical view of these vortical structures 
generated in the near wall region by the alternating no-slip/free-shear boundary conditions 
and their dependency on the solid fraction. 
The effect of the solid fraction at constant periodicity is peculiar since the structures
are anchored to the perfect-slip/no-slip boundary and their dimensions depend on the
grooves width. 
Two pairs of eddies are present over the interface. The largest one develops far from the wall whilst 
the smallest pair is located very close to the wall and counter-rotate with respect to the large ones. This qualitative behaviour 
is independent of the solid fraction and of the width of the grooves. 
The largest vortices are coherent with the mean radial velocity configuration observed in figure~\ref{fig:ur_phase}.
Over the grooves, the largest eddies move fluid with high velocity 
from the bulk of the flow towards the free-shear region, contributing to the increment in mean slip velocity $\langle u_s \rangle$.
The mean slip velocity increases with the width of the grooves since, due to a larger extension in the bulk of the flow, 
these structures interact with higher velocity regions in cases D50 and E50.
Furthermore, the vortex motion is also responsible for the increase in wall shear stress, shown in figure~\ref{fig:uz_ph}.
These structures move the high velocity fluid over the liquid/gas interface towards
the solid wall (no slip) region, increasing the shear stress. 
It is important to note that these vortical structures are permanent, unlike the hairpin vortices in classical smooth pipe turbulence 
which are instantaneous intermittent structures. Another difference with respect to smooth pipe flow is that  
the negative radial velocity produces an increase in local shear stress since the positive radial velocity occurs in the
regions where shear-free stress is enforced.
Comparing cases D50 and D75 (panels (a) and (c) respectively), it is clear that these structures are anchored with
the wall-groove interfaces and their radial dimensions are related to the width of the grooves. Figure~\ref{fig:core_dist} 
shows the distance between the centres of the two mean vortices represented in figure~\ref{fig:phase_struct} as a function of the solid
fraction normalised with the groove width.
The figure shows that the distance between vortices is strongly related to the groove width independently of the solid fraction, and these vortical structures 
are anchored to the no-slip/shear-free interfaces. 
\begin{figure}[h]
\centering
\includegraphics[width=.45\textwidth]{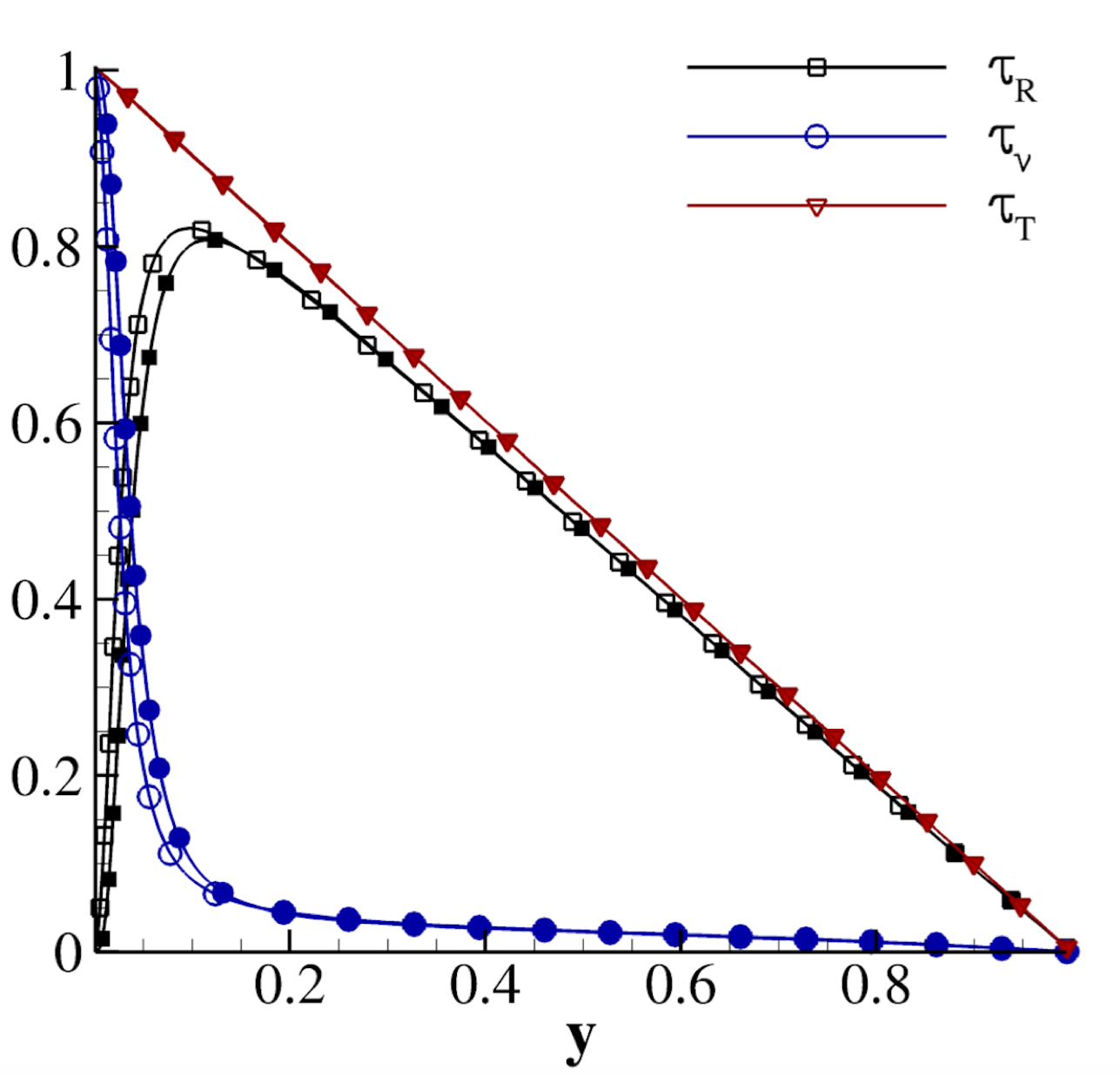}
\caption{\label{fig:stress_all} Balance of the Reynolds stress $\tau_R= \langle u'_r u'_z\rangle$, 
viscous stress $\tau_\nu=1/{\rm Re}\,\partial \langle u_z\rangle/\partial r$ and of the total stress 
$\tau_T=\tau_R+\tau_\nu$ normalised with the wall shear stress 
$\tau_w=\left.1/{\rm Re}\,\partial \langle u_z\rangle/\partial r\right|_{r=1}$.  	
Closed and open symbols represent the REF and D50 cases respectively.
}
\end{figure}
\begin{figure}[h]
\centering
\includegraphics[width=.9\textwidth]{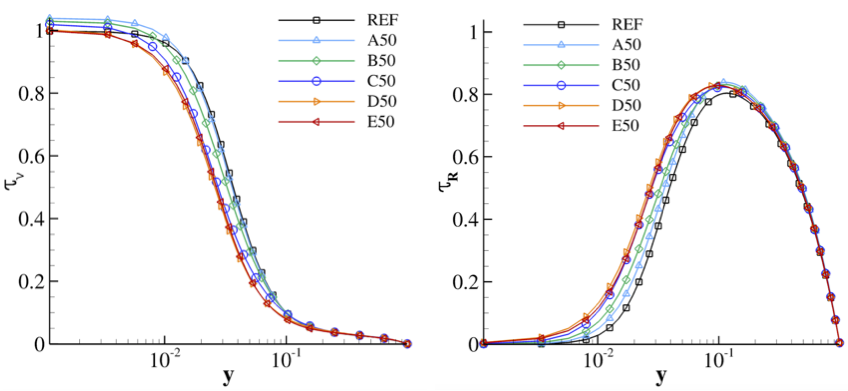}
\caption{\label{fig:visc_stress} Left panel: radial profiles of the viscous stress 
$\tau_\nu=1/{\rm Re}\,\partial \langle u_z\rangle/\partial r$ normalised 
with the wall shear stress $\tau_w=\left.1/{\rm Re}\,\partial \langle u_z\rangle/\partial r\right|_{r=1}$ for different
groove dimensions. Right panel: radial profiles of the Reynolds stress 
$\tau_R= \langle u'_r u'_z\rangle$ normalised 
with the wall shear stress $\tau_w=\left.1/{\rm Re}\,\partial \langle u_z\rangle/\partial r\right|_{r=1}$ for different
grooves dimensions.
}
\end{figure}
\subsection{Overall momentum balance}
\label{sec:mom_bal}

The flow rate increase due to the no-slip/free-shear alternating boundary conditions can be expanded on. 
Two different phenomena are induced by the permanent near wall structures observed in figure~\ref{fig:phase_struct};  
the generation of a mean slip velocity at the liquid/gas interface and the increase in local shear-stress at the solid wall. 
Each influences the overall flow rate. In order to distinguish the different 
contributions, the momentum balance developed by~\citet{fuka_bil} is generalised for the present cases. 
For this purpose, the Reynolds-averaged momentum equation in the streamwise direction is necessary,
\begin{align}
\left.-\frac{\text{d}p}{\text{d}z}\right|_0 + \frac{1}{r} \frac{\text{d}}
{\text{d}r} \left[ r \left(\frac{1}{\rm{Re}} \frac{\text{d}\langle u_z\rangle}{\text{d}r} + \tau_R \right) \right]=0\, ,
\label{eq:fukag}
\end{align}
where $\tau_R = - \langle u'_r u'_z \rangle$ is the shear component of the Reynolds stress tensor.
Following the classical procedure, it is possible to find that the global stress $\tau_T$, which is the sum
of the viscous and the shear Reynolds stresses, varies linearly with the pipe radius. Figure~\ref{fig:stress_all}
shows the viscous and shear Reynolds stresses against radial distance, together with their sum. The 
comparison between REF and D50 cases shows that the boundary condition effects 
are limited to the region close to the wall, shown by a shear Reynolds stress increase and the decrease of 
the viscous stresses for the superhydrophobic case D50. Figure~\ref{fig:visc_stress} shows both the viscous 
stresses, panel (a), and the shear Reynolds stresses, panel (b), for the remaining cases, confirming the
observations. 
On the other hand, following the procedure reported in~\citet{fuka_bil}, equation~\eqref{eq:fukag} can be recast as 
\begin{equation}
R^2 U_b - R^2 \langle u_s\rangle 
- \rm{Re} \int_{0}^{R}r^2 \, \tau_R \left( r \right) \,dr = - \frac{ \rm{Re}}{8} \left. \frac{\text{d}p}{\text{d}z} \right|_0 R^4 \, . 
\label{eq:fkag_s}
\end{equation}
Equation~\eqref{eq:fkag_s} states that the pressure gradient that sustains the flow (right-hand-side) is balanced by 
the global flow rate (first term on the left hand side), by the flow rate due to the slip velocity (second term) and by the Reynolds stress (third term). 
In standard turbulent pipe flow $\langle u_s\rangle=0$, and the equation simplifies to the one obtained in~\citet{fuka_bil}. 

\begin{figure}[h]
\centering
\includegraphics[width=\textwidth]{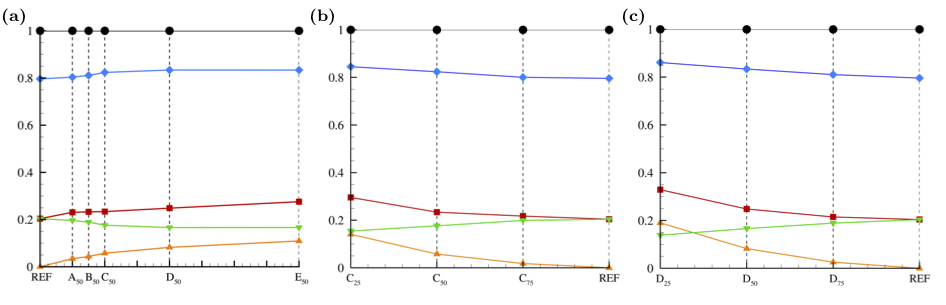}
\caption{\label{fig:fuk} Balance of equation~\eqref{eq:fkag_s} depending on: the groove  
dimension at fixed solid fraction $\Phi_S=0.5$ in the left panel and the solid fraction for different periodicity in the middle
and right panels. Colours indicate different terms on the left-hand-side of 
equation~\eqref{eq:fkag_s} which are normalised with the right-hand-side term. The first term is denoted in 
red, the second in the orange, the subtraction of the first two terms is in green, the third 
term in blue and the normalised term in the right-hand-side (which has to be equal to one) is in black.
}
\end{figure}

Figure~\ref{fig:fuk} shows the terms in equation~\eqref{eq:fkag_s} normalised with the 
right-hand-side term. The overall flow rate is drawn with red squares, the flow rate related to the 
slip velocity is drawn with orange triangles (the sum of the first and second terms is reported with 
green triangles), the third term related with the Reynolds stress is 
drawn with blue diamonds and the overall sum is drawn with black circles and has to be equal one.
The present simulations are performed at constant pressure gradient, fixing the right-hand-side of  
equation~\eqref{eq:fkag_s}. The standard configuration of flow in a turbulent pipe flow presents a lower 
flow rate with respect to laminar flow due to the positive contribution of the Reynolds stress integral 
(third term in the left-hand-side of equation~\eqref{eq:fkag_s}). 
Consider the sum of the first and second term in equation~\eqref{eq:fkag_s}, representing 
the flow rate associated with $U_b-\langle u_s\rangle$, i.e. the integral of the profiles in figure~\ref{fig:uz}
panels (b), (d) and (f). The trend of this combination depends on the shear stress $\tau_R$ contribution. 
The increase in shear stress results in a decrease in the sum of the first two terms, see 
green triangles in figure~\ref{fig:fuk}, which is consistent with the profiles in figure~\ref{fig:uz}. 
The second contribution of this sum is observed to increase with the groove width, see orange triangles, 
hence the overall flow rate has to increase consistently, see red squares. These observations are in agreement  
with the profiles in the figure~\ref{fig:uz}. Changing the solid fraction results in the same trends. 
Increasing the solid fraction results in a decrease in shear stresses and therefore the overall flow rate decreases
following the decrease in slip velocity. 

\section{Conclusions}

The effects of superhydrophobic surfaces on turbulent pipe flow have been studied.
Whilst there are several works concerning planar channel flow in the literature, we 
address the turbulence modulation induced by these surfaces in turbulent pipe flow
at friction Reynolds number higher than that available in the literature. The effect of 
the periodicity length and the solid fraction on turbulent structures and fluctuations is addressed.

Two phenomena are observed when fixing the solid fraction and increasing the periodicity length. Firstly, the mean slip 
velocity at the wall increases with increasing periodicity length, therefore increasing the flow rate. Secondly, the superhydrophobic 
boundary conditions modulate the turbulent structures and this results in a decrease in flow rate with the increase in periodicity length. 
Between the two concurrent effects, the increase in mean slip velocity overwhelms the turbulence modulation effect which 
mitigates the positive drag reducing effects of the former. 
The boundary conditions directly influence the flow rate since all the simulations are performed at the same fixed pressure gradient.
The flow rate increase can therefore be interpreted as a drag decrease. 
On the other hand, when fixing the periodicity length and increasing the solid fraction, the mean slip velocity decreases. 
This results in a decrease in flow rate. The turbulence modification becomes negligible as the solid fraction increases.

The no-slip/shear-free alternating boundary conditions also induce some peculiar and persistent mean 
vortical structures that are anchored to the no-slip/shear-free interfaces. The distance between two close vortical structures 
strongly depends on the width of the grooves and is independent of the solid fraction. 
The vortices induce radial velocities towards the liquid/gas interfaces and away from the walls. 
These structures can induce undesired motions to the liquid/gas interfaces, which may eventually wet  
the wall asperities and result in additional drag with respect to the smooth walls. 

Finally, the overall balance of the axial momentum describes the different contributions to the flow rate. The 
no-slip/shear-free alternating boundary conditions produce an increase in Reynolds stresses, which correspond 
to a drag increase due to the alteration of turbulence. This effect is overwhelmed by the effect of the mean 
slip velocity which produces an increase in flow rate, and therefore a decrease in drag. This drag reduction is the 
ultimate positive effect of applying superhydrophobic textures to walls.

\section*{Acknowledgements}
The research received funding from the European Research Council under the 
European Union's Seventh Framework Programme (FP7/2007-2013)/ERC grant agreement 
no. [339446]. Support from  PRACE,  projects FP7 RI-283493 and grant 
no. 2014112647, is acknowledged.
\bibliography{biblio}

\begin{thebibliography}{35}%
\makeatletter
\providecommand \@ifxundefined [1]{%
 \@ifx{#1\undefined}
}%
\providecommand \@ifnum [1]{%
 \ifnum #1\expandafter \@firstoftwo
 \else \expandafter \@secondoftwo
 \fi
}%
\providecommand \@ifx [1]{%
 \ifx #1\expandafter \@firstoftwo
 \else \expandafter \@secondoftwo
 \fi
}%
\providecommand \natexlab [1]{#1}%
\providecommand \enquote  [1]{``#1''}%
\providecommand \bibnamefont  [1]{#1}%
\providecommand \bibfnamefont [1]{#1}%
\providecommand \citenamefont [1]{#1}%
\providecommand \href@noop [0]{\@secondoftwo}%
\providecommand \href [0]{\begingroup \@sanitize@url \@href}%
\providecommand \@href[1]{\@@startlink{#1}\@@href}%
\providecommand \@@href[1]{\endgroup#1\@@endlink}%
\providecommand \@sanitize@url [0]{\catcode `\\12\catcode `\$12\catcode
  `\&12\catcode `\#12\catcode `\^12\catcode `\_12\catcode `\%12\relax}%
\providecommand \@@startlink[1]{}%
\providecommand \@@endlink[0]{}%
\providecommand \url  [0]{\begingroup\@sanitize@url \@url }%
\providecommand \@url [1]{\endgroup\@href {#1}{\urlprefix }}%
\providecommand \urlprefix  [0]{URL }%
\providecommand \Eprint [0]{\href }%
\providecommand \doibase [0]{http://dx.doi.org/}%
\providecommand \selectlanguage [0]{\@gobble}%
\providecommand \bibinfo  [0]{\@secondoftwo}%
\providecommand \bibfield  [0]{\@secondoftwo}%
\providecommand \translation [1]{[#1]}%
\providecommand \BibitemOpen [0]{}%
\providecommand \bibitemStop [0]{}%
\providecommand \bibitemNoStop [0]{.\EOS\space}%
\providecommand \EOS [0]{\spacefactor3000\relax}%
\providecommand \BibitemShut  [1]{\csname bibitem#1\endcsname}%
\let\auto@bib@innerbib\@empty
\bibitem [{\citenamefont {Frohnapfel}, \citenamefont {Hasegawa},\ and\
  \citenamefont {Quadrio}(2012)}]{frohn}%
  \BibitemOpen
  \bibfield  {author} {\bibinfo {author} {\bibfnamefont {B.}~\bibnamefont
  {Frohnapfel}}, \bibinfo {author} {\bibfnamefont {Y.}~\bibnamefont
  {Hasegawa}}, \ and\ \bibinfo {author} {\bibfnamefont {M.}~\bibnamefont
  {Quadrio}},\ }\bibfield  {title} {\enquote {\bibinfo {title} {Money versus
  time: evaluation of flow control in terms of energy consumption and
  convenience},}\ }\href@noop {} {\bibfield  {journal} {\bibinfo  {journal}
  {Journal of Fluid Mechanics}\ }\textbf {\bibinfo {volume} {700}},\ \bibinfo
  {pages} {406--418} (\bibinfo {year} {2012})}\BibitemShut {NoStop}%
\bibitem [{\citenamefont {White}\ and\ \citenamefont {Mungal}(2008)}]{white}%
  \BibitemOpen
  \bibfield  {author} {\bibinfo {author} {\bibfnamefont {C.~M.}\ \bibnamefont
  {White}}\ and\ \bibinfo {author} {\bibfnamefont {M.~G.}\ \bibnamefont
  {Mungal}},\ }\bibfield  {title} {\enquote {\bibinfo {title} {Mechanics and
  prediction of turbulent drag reduction with polymer additives},}\ }\href@noop
  {} {\bibfield  {journal} {\bibinfo  {journal} {Annual Review of Fluid
  Mechanics}\ }\textbf {\bibinfo {volume} {40}},\ \bibinfo {pages} {235--256}
  (\bibinfo {year} {2008})}\BibitemShut {NoStop}%
\bibitem [{\citenamefont {Ceccio}(2010)}]{ceccio}%
  \BibitemOpen
  \bibfield  {author} {\bibinfo {author} {\bibfnamefont {S.~L.}\ \bibnamefont
  {Ceccio}},\ }\bibfield  {title} {\enquote {\bibinfo {title} {Friction drag
  reduction of external flows with bubble and gas injection},}\ }\href@noop {}
  {\bibfield  {journal} {\bibinfo  {journal} {Annual Review of Fluid
  Mechanics}\ }\textbf {\bibinfo {volume} {42}},\ \bibinfo {pages} {183--203}
  (\bibinfo {year} {2010})}\BibitemShut {NoStop}%
\bibitem [{\citenamefont {Spalart}\ and\ \citenamefont
  {McLean}(2011)}]{spalart}%
  \BibitemOpen
  \bibfield  {author} {\bibinfo {author} {\bibfnamefont {P.~R.}\ \bibnamefont
  {Spalart}}\ and\ \bibinfo {author} {\bibfnamefont {J.~D.}\ \bibnamefont
  {McLean}},\ }\bibfield  {title} {\enquote {\bibinfo {title} {Drag reduction:
  enticing turbulence, and then an industry},}\ }\href@noop {} {\bibfield
  {journal} {\bibinfo  {journal} {Philosophical Transactions of the Royal
  Society of London A: Mathematical, Physical and Engineering Sciences}\
  }\textbf {\bibinfo {volume} {369}},\ \bibinfo {pages} {1556--1569} (\bibinfo
  {year} {2011})}\BibitemShut {NoStop}%
\bibitem [{\citenamefont {Lafuma}\ and\ \citenamefont
  {Qu\'er\'e}(2003)}]{lafuma}%
  \BibitemOpen
  \bibfield  {author} {\bibinfo {author} {\bibfnamefont {A.}~\bibnamefont
  {Lafuma}}\ and\ \bibinfo {author} {\bibfnamefont {D.}~\bibnamefont
  {Qu\'er\'e}},\ }\bibfield  {title} {\enquote {\bibinfo {title}
  {Superhydrophobic states},}\ }\href@noop {} {\bibfield  {journal} {\bibinfo
  {journal} {Nature Materials}\ }\textbf {\bibinfo {volume} {2}},\ \bibinfo
  {pages} {457--460} (\bibinfo {year} {2003})}\BibitemShut {NoStop}%
\bibitem [{\citenamefont {Wenzel}(1936)}]{wenzel}%
  \BibitemOpen
  \bibfield  {author} {\bibinfo {author} {\bibfnamefont {R.~N.}\ \bibnamefont
  {Wenzel}},\ }\bibfield  {title} {\enquote {\bibinfo {title} {Resistance of
  solid surfaces to wetting by water},}\ }\href@noop {} {\bibfield  {journal}
  {\bibinfo  {journal} {Industrial \& Engineering Chemistry}\ }\textbf
  {\bibinfo {volume} {28}},\ \bibinfo {pages} {988--994} (\bibinfo {year}
  {1936})}\BibitemShut {NoStop}%
\bibitem [{\citenamefont {Cassie}\ and\ \citenamefont {Baxter}(1944)}]{cassie}%
  \BibitemOpen
  \bibfield  {author} {\bibinfo {author} {\bibfnamefont {A.~B.~D.}\
  \bibnamefont {Cassie}}\ and\ \bibinfo {author} {\bibfnamefont
  {S.}~\bibnamefont {Baxter}},\ }\bibfield  {title} {\enquote {\bibinfo {title}
  {Wettability of porous surfaces},}\ }\href@noop {} {\bibfield  {journal}
  {\bibinfo  {journal} {Trans. Faraday Soc.}\ }\textbf {\bibinfo {volume}
  {40}},\ \bibinfo {pages} {546--551} (\bibinfo {year} {1944})}\BibitemShut
  {NoStop}%
\bibitem [{\citenamefont {Ou}, \citenamefont {Perot},\ and\ \citenamefont
  {Rothstein}(2004)}]{ou}%
  \BibitemOpen
  \bibfield  {author} {\bibinfo {author} {\bibfnamefont {J.}~\bibnamefont
  {Ou}}, \bibinfo {author} {\bibfnamefont {B.}~\bibnamefont {Perot}}, \ and\
  \bibinfo {author} {\bibfnamefont {J.}~\bibnamefont {Rothstein}},\ }\bibfield
  {title} {\enquote {\bibinfo {title} {Laminar drag reduction in microchannels
  using ultrahydrophobic surfaces},}\ }\href@noop {} {\bibfield  {journal}
  {\bibinfo  {journal} {Physics of Fluids}\ }\textbf {\bibinfo {volume} {16}},\
  \bibinfo {pages} {4635--4643} (\bibinfo {year} {2004})}\BibitemShut {NoStop}%
\bibitem [{\citenamefont {Rothstein}(2010)}]{roth}%
  \BibitemOpen
  \bibfield  {author} {\bibinfo {author} {\bibfnamefont {J.~P.}\ \bibnamefont
  {Rothstein}},\ }\bibfield  {title} {\enquote {\bibinfo {title} {Slip on
  superhydrophobic surfaces},}\ }\href@noop {} {\bibfield  {journal} {\bibinfo
  {journal} {Annual Review of Fluid Mechanics}\ }\textbf {\bibinfo {volume}
  {42}},\ \bibinfo {pages} {89--109} (\bibinfo {year} {2010})}\BibitemShut
  {NoStop}%
\bibitem [{\citenamefont {Pimponi}\ \emph {et~al.}(2014)\citenamefont
  {Pimponi}, \citenamefont {Chinappi}, \citenamefont {Gualtieri},\ and\
  \citenamefont {Casciola}}]{pimponi}%
  \BibitemOpen
  \bibfield  {author} {\bibinfo {author} {\bibfnamefont {D.}~\bibnamefont
  {Pimponi}}, \bibinfo {author} {\bibfnamefont {M.}~\bibnamefont {Chinappi}},
  \bibinfo {author} {\bibfnamefont {P.}~\bibnamefont {Gualtieri}}, \ and\
  \bibinfo {author} {\bibfnamefont {C.~M.}\ \bibnamefont {Casciola}},\
  }\bibfield  {title} {\enquote {\bibinfo {title} {Mobility tensor of a sphere
  moving on a superhydrophobic wall: application to particle separation},}\
  }\href@noop {} {\bibfield  {journal} {\bibinfo  {journal} {Microfluidics and
  nanofluidics}\ }\textbf {\bibinfo {volume} {16}},\ \bibinfo {pages}
  {571--585} (\bibinfo {year} {2014})}\BibitemShut {NoStop}%
\bibitem [{\citenamefont {Min}\ and\ \citenamefont {Kim}(2004)}]{min}%
  \BibitemOpen
  \bibfield  {author} {\bibinfo {author} {\bibfnamefont {T.}~\bibnamefont
  {Min}}\ and\ \bibinfo {author} {\bibfnamefont {J.}~\bibnamefont {Kim}},\
  }\bibfield  {title} {\enquote {\bibinfo {title} {Effects of hydrophobic
  surface on skin-friction drag},}\ }\href@noop {} {\bibfield  {journal}
  {\bibinfo  {journal} {Physics of Fluids}\ }\textbf {\bibinfo {volume} {16}},\
  \bibinfo {pages} {L55--L58} (\bibinfo {year} {2004})}\BibitemShut {NoStop}%
\bibitem [{\citenamefont {Busse}\ and\ \citenamefont {Sandham}(2012)}]{busse}%
  \BibitemOpen
  \bibfield  {author} {\bibinfo {author} {\bibfnamefont {A.}~\bibnamefont
  {Busse}}\ and\ \bibinfo {author} {\bibfnamefont {N.}~\bibnamefont
  {Sandham}},\ }\bibfield  {title} {\enquote {\bibinfo {title} {Influence of an
  anisotropic slip-length boundary condition on turbulent channel flow},}\
  }\href@noop {} {\bibfield  {journal} {\bibinfo  {journal} {Physics of
  Fluids}\ }\textbf {\bibinfo {volume} {24}},\ \bibinfo {pages} {055111}
  (\bibinfo {year} {2012})}\BibitemShut {NoStop}%
\bibitem [{\citenamefont {Martell}, \citenamefont {Rothstein},\ and\
  \citenamefont {Perot}(2010)}]{martell}%
  \BibitemOpen
  \bibfield  {author} {\bibinfo {author} {\bibfnamefont {M.~B.}\ \bibnamefont
  {Martell}}, \bibinfo {author} {\bibfnamefont {J.~P.}\ \bibnamefont
  {Rothstein}}, \ and\ \bibinfo {author} {\bibfnamefont {J.~B.}\ \bibnamefont
  {Perot}},\ }\bibfield  {title} {\enquote {\bibinfo {title} {An analysis of
  superhydrophobic turbulent drag reduction mechanisms using direct numerical
  simulation},}\ }\href@noop {} {\bibfield  {journal} {\bibinfo  {journal}
  {Physics of Fluids}\ }\textbf {\bibinfo {volume} {22}},\ \bibinfo {eid}
  {065102} (\bibinfo {year} {2010})}\BibitemShut {NoStop}%
\bibitem [{\citenamefont {Fukagata}, \citenamefont {Kasagi},\ and\
  \citenamefont {Koumoutsakos}(2006)}]{fukagata}%
  \BibitemOpen
  \bibfield  {author} {\bibinfo {author} {\bibfnamefont {K.}~\bibnamefont
  {Fukagata}}, \bibinfo {author} {\bibfnamefont {N.}~\bibnamefont {Kasagi}}, \
  and\ \bibinfo {author} {\bibfnamefont {P.}~\bibnamefont {Koumoutsakos}},\
  }\bibfield  {title} {\enquote {\bibinfo {title} {A theoretical prediction of
  friction drag reduction in turbulent flow by superhydrophobic surfaces},}\
  }\href@noop {} {\bibfield  {journal} {\bibinfo  {journal} {Physics of
  Fluids}\ }\textbf {\bibinfo {volume} {18}},\ \bibinfo {eid} {051703}
  (\bibinfo {year} {2006})}\BibitemShut {NoStop}%
\bibitem [{\citenamefont {Park}, \citenamefont {Park},\ and\ \citenamefont
  {Kim}(2013)}]{park}%
  \BibitemOpen
  \bibfield  {author} {\bibinfo {author} {\bibfnamefont {H.}~\bibnamefont
  {Park}}, \bibinfo {author} {\bibfnamefont {H.}~\bibnamefont {Park}}, \ and\
  \bibinfo {author} {\bibfnamefont {J.}~\bibnamefont {Kim}},\ }\bibfield
  {title} {\enquote {\bibinfo {title} {A numerical study of the effects of
  superhydrophobic surface on skin-friction drag in turbulent channel flow},}\
  }\href@noop {} {\bibfield  {journal} {\bibinfo  {journal} {Physics of
  Fluids}\ }\textbf {\bibinfo {volume} {25}},\ \bibinfo {eid} {110815}
  (\bibinfo {year} {2013})}\BibitemShut {NoStop}%
\bibitem [{\citenamefont {Rastegari}\ and\ \citenamefont
  {Akhavan}(2015)}]{rastegari2015}%
  \BibitemOpen
  \bibfield  {author} {\bibinfo {author} {\bibfnamefont {A.}~\bibnamefont
  {Rastegari}}\ and\ \bibinfo {author} {\bibfnamefont {R.}~\bibnamefont
  {Akhavan}},\ }\bibfield  {title} {\enquote {\bibinfo {title} {On the
  mechanism of turbulent drag reduction with super-hydrophobic surfaces},}\
  }\href@noop {} {\bibfield  {journal} {\bibinfo  {journal} {Journal of Fluid
  Mechanics}\ }\textbf {\bibinfo {volume} {773}} (\bibinfo {year}
  {2015})}\BibitemShut {NoStop}%
\bibitem [{\citenamefont {T{\"u}rk}\ \emph {et~al.}(2014)\citenamefont
  {T{\"u}rk}, \citenamefont {Daschiel}, \citenamefont {Stroh}, \citenamefont
  {Hasegawa},\ and\ \citenamefont {Frohnapfel}}]{turk}%
  \BibitemOpen
  \bibfield  {author} {\bibinfo {author} {\bibfnamefont {S.}~\bibnamefont
  {T{\"u}rk}}, \bibinfo {author} {\bibfnamefont {G.}~\bibnamefont {Daschiel}},
  \bibinfo {author} {\bibfnamefont {A.}~\bibnamefont {Stroh}}, \bibinfo
  {author} {\bibfnamefont {Y.}~\bibnamefont {Hasegawa}}, \ and\ \bibinfo
  {author} {\bibfnamefont {B.}~\bibnamefont {Frohnapfel}},\ }\bibfield  {title}
  {\enquote {\bibinfo {title} {Turbulent flow over superhydrophobic surfaces
  with streamwise grooves},}\ }\href@noop {} {\bibfield  {journal} {\bibinfo
  {journal} {Journal of Fluid Mechanics}\ }\textbf {\bibinfo {volume} {747}},\
  \bibinfo {pages} {186--217} (\bibinfo {year} {2014})}\BibitemShut {NoStop}%
\bibitem [{\citenamefont {Stroh}\ \emph {et~al.}()\citenamefont {Stroh},
  \citenamefont {Hasegawa}, \citenamefont {J},\ and\ \citenamefont
  {Frohnapfel}}]{stroh}%
  \BibitemOpen
  \bibfield  {author} {\bibinfo {author} {\bibfnamefont {A.}~\bibnamefont
  {Stroh}}, \bibinfo {author} {\bibfnamefont {Y.}~\bibnamefont {Hasegawa}},
  \bibinfo {author} {\bibfnamefont {K.}~\bibnamefont {J}}, \ and\ \bibinfo
  {author} {\bibfnamefont {B.}~\bibnamefont {Frohnapfel}},\ }\bibfield  {title}
  {\enquote {\bibinfo {title} {Wave-length-dependent rearrangement of secondary
  vortices over superhydrophobic surfaces with streamwise grooves},}\ }in\
  \href@noop {} {\emph {\bibinfo {booktitle} {The 9th Symposium on Turbulence
  and Shear Flow Phenomena (TSFP-9)(June 30th--July 3rd, Melbourne,
  Austalia)}}},\ pp.\ \bibinfo {pages} {9A--1}\BibitemShut {NoStop}%
\bibitem [{\citenamefont {Im}\ and\ \citenamefont
  {Lee}(2017)}]{im2017comparison}%
  \BibitemOpen
  \bibfield  {author} {\bibinfo {author} {\bibfnamefont {H.~J.}\ \bibnamefont
  {Im}}\ and\ \bibinfo {author} {\bibfnamefont {J.~H.}\ \bibnamefont {Lee}},\
  }\bibfield  {title} {\enquote {\bibinfo {title} {Comparison of
  superhydrophobic drag reduction between turbulent pipe and channel flows},}\
  }\href@noop {} {\bibfield  {journal} {\bibinfo  {journal} {Physics of
  Fluids}\ }\textbf {\bibinfo {volume} {29}},\ \bibinfo {pages} {095101}
  (\bibinfo {year} {2017})}\BibitemShut {NoStop}%
\bibitem [{\citenamefont {Tian}\ \emph {et~al.}(2015)\citenamefont {Tian},
  \citenamefont {Zhang}, \citenamefont {Jiang},\ and\ \citenamefont
  {Yao}}]{tian}%
  \BibitemOpen
  \bibfield  {author} {\bibinfo {author} {\bibfnamefont {H.}~\bibnamefont
  {Tian}}, \bibinfo {author} {\bibfnamefont {J.}~\bibnamefont {Zhang}},
  \bibinfo {author} {\bibfnamefont {N.}~\bibnamefont {Jiang}}, \ and\ \bibinfo
  {author} {\bibfnamefont {Z.}~\bibnamefont {Yao}},\ }\bibfield  {title}
  {\enquote {\bibinfo {title} {Effect of hierarchical structured
  superhydrophobic surfaces on coherent structures in turbulent channel
  flow},}\ }\href@noop {} {\bibfield  {journal} {\bibinfo  {journal}
  {Experimental Thermal and Fluid Science}\ }\textbf {\bibinfo {volume} {69}},\
  \bibinfo {pages} {27--37} (\bibinfo {year} {2015})}\BibitemShut {NoStop}%
\bibitem [{\citenamefont {Henoch}\ \emph {et~al.}(2006)\citenamefont {Henoch},
  \citenamefont {Krupenkin}, \citenamefont {Kolodner}, \citenamefont {Taylor},
  \citenamefont {Hodes}, \citenamefont {Lyons}, \citenamefont {Peguero},\ and\
  \citenamefont {Breuer}}]{henoch}%
  \BibitemOpen
  \bibfield  {author} {\bibinfo {author} {\bibfnamefont {C.}~\bibnamefont
  {Henoch}}, \bibinfo {author} {\bibfnamefont {T.~N.}\ \bibnamefont
  {Krupenkin}}, \bibinfo {author} {\bibfnamefont {P.}~\bibnamefont {Kolodner}},
  \bibinfo {author} {\bibfnamefont {J.~A.}\ \bibnamefont {Taylor}}, \bibinfo
  {author} {\bibfnamefont {M.~S.}\ \bibnamefont {Hodes}}, \bibinfo {author}
  {\bibfnamefont {A.~M.}\ \bibnamefont {Lyons}}, \bibinfo {author}
  {\bibfnamefont {C.}~\bibnamefont {Peguero}}, \ and\ \bibinfo {author}
  {\bibfnamefont {K.}~\bibnamefont {Breuer}},\ }\bibfield  {title} {\enquote
  {\bibinfo {title} {Turbulent drag reduction using superhydrophobic
  surfaces},}\ }in\ \href@noop {} {\emph {\bibinfo {booktitle} {3rd AIAA Flow
  Control Conference}}}\ (\bibinfo {year} {2006})\ pp.\ \bibinfo {pages}
  {5--8}\BibitemShut {NoStop}%
\bibitem [{\citenamefont {Daniello}, \citenamefont {Waterhouse},\ and\
  \citenamefont {Rothstein}(2009)}]{daniello}%
  \BibitemOpen
  \bibfield  {author} {\bibinfo {author} {\bibfnamefont {R.~J.}\ \bibnamefont
  {Daniello}}, \bibinfo {author} {\bibfnamefont {N.~E.}\ \bibnamefont
  {Waterhouse}}, \ and\ \bibinfo {author} {\bibfnamefont {J.~P.}\ \bibnamefont
  {Rothstein}},\ }\bibfield  {title} {\enquote {\bibinfo {title} {Drag
  reduction in turbulent flows over superhydrophobic surfaces},}\ }\href@noop
  {} {\bibfield  {journal} {\bibinfo  {journal} {Physics of Fluids}\ }\textbf
  {\bibinfo {volume} {21}},\ \bibinfo {eid} {085103} (\bibinfo {year}
  {2009})}\BibitemShut {NoStop}%
\bibitem [{\citenamefont {Amabili}\ \emph {et~al.}(2016)\citenamefont
  {Amabili}, \citenamefont {Lisi}, \citenamefont {Giacomello},\ and\
  \citenamefont {Casciola}}]{amabili}%
  \BibitemOpen
  \bibfield  {author} {\bibinfo {author} {\bibfnamefont {M.}~\bibnamefont
  {Amabili}}, \bibinfo {author} {\bibfnamefont {E.}~\bibnamefont {Lisi}},
  \bibinfo {author} {\bibfnamefont {A.}~\bibnamefont {Giacomello}}, \ and\
  \bibinfo {author} {\bibfnamefont {C.}~\bibnamefont {Casciola}},\ }\bibfield
  {title} {\enquote {\bibinfo {title} {Wetting and cavitation pathways on
  nanodecorated surfaces},}\ }\href@noop {} {\bibfield  {journal} {\bibinfo
  {journal} {Soft matter}\ }\textbf {\bibinfo {volume} {12}},\ \bibinfo {pages}
  {3046--3055} (\bibinfo {year} {2016})}\BibitemShut {NoStop}%
\bibitem [{\citenamefont {Aljallis}\ \emph {et~al.}(2013)\citenamefont
  {Aljallis}, \citenamefont {Sarshar}, \citenamefont {Datla}, \citenamefont
  {Sikka}, \citenamefont {Jones},\ and\ \citenamefont {Choi}}]{alj}%
  \BibitemOpen
  \bibfield  {author} {\bibinfo {author} {\bibfnamefont {E.}~\bibnamefont
  {Aljallis}}, \bibinfo {author} {\bibfnamefont {M.~A.}\ \bibnamefont
  {Sarshar}}, \bibinfo {author} {\bibfnamefont {R.}~\bibnamefont {Datla}},
  \bibinfo {author} {\bibfnamefont {V.}~\bibnamefont {Sikka}}, \bibinfo
  {author} {\bibfnamefont {A.}~\bibnamefont {Jones}}, \ and\ \bibinfo {author}
  {\bibfnamefont {C.-H.}\ \bibnamefont {Choi}},\ }\bibfield  {title} {\enquote
  {\bibinfo {title} {Experimental study of skin friction drag reduction on
  superhydrophobic flat plates in high reynolds number boundary layer flow},}\
  }\href@noop {} {\bibfield  {journal} {\bibinfo  {journal} {Physics of
  fluids}\ }\textbf {\bibinfo {volume} {25}},\ \bibinfo {pages} {025103}
  (\bibinfo {year} {2013})}\BibitemShut {NoStop}%
\bibitem [{\citenamefont {Battista}, \citenamefont {Picano},\ and\
  \citenamefont {Casciola}(2014)}]{battista2014turbulent}%
  \BibitemOpen
  \bibfield  {author} {\bibinfo {author} {\bibfnamefont {F.}~\bibnamefont
  {Battista}}, \bibinfo {author} {\bibfnamefont {F.}~\bibnamefont {Picano}}, \
  and\ \bibinfo {author} {\bibfnamefont {C.}~\bibnamefont {Casciola}},\
  }\bibfield  {title} {\enquote {\bibinfo {title} {Turbulent mixing of a
  slightly supercritical van der waals fluid at low-mach number},}\ }\href@noop
  {} {\bibfield  {journal} {\bibinfo  {journal} {Physics of Fluids}\ }\textbf
  {\bibinfo {volume} {26}},\ \bibinfo {pages} {055101} (\bibinfo {year}
  {2014})}\BibitemShut {NoStop}%
\bibitem [{\citenamefont {Battista}, \citenamefont {Troiani},\ and\
  \citenamefont {Picano}(2015)}]{battista2015fractal}%
  \BibitemOpen
  \bibfield  {author} {\bibinfo {author} {\bibfnamefont {F.}~\bibnamefont
  {Battista}}, \bibinfo {author} {\bibfnamefont {G.}~\bibnamefont {Troiani}}, \
  and\ \bibinfo {author} {\bibfnamefont {F.}~\bibnamefont {Picano}},\
  }\bibfield  {title} {\enquote {\bibinfo {title} {Fractal scaling of turbulent
  premixed flame fronts: Application to les},}\ }\href@noop {} {\bibfield
  {journal} {\bibinfo  {journal} {International Journal of Heat and Fluid
  Flow}\ }\textbf {\bibinfo {volume} {51}},\ \bibinfo {pages} {78--87}
  (\bibinfo {year} {2015})}\BibitemShut {NoStop}%
\bibitem [{\citenamefont {Rocco}\ \emph {et~al.}(2015)\citenamefont {Rocco},
  \citenamefont {Battista}, \citenamefont {Picano}, \citenamefont {Troiani},\
  and\ \citenamefont {Casciola}}]{rocco2015curvature}%
  \BibitemOpen
  \bibfield  {author} {\bibinfo {author} {\bibfnamefont {G.}~\bibnamefont
  {Rocco}}, \bibinfo {author} {\bibfnamefont {F.}~\bibnamefont {Battista}},
  \bibinfo {author} {\bibfnamefont {F.}~\bibnamefont {Picano}}, \bibinfo
  {author} {\bibfnamefont {G.}~\bibnamefont {Troiani}}, \ and\ \bibinfo
  {author} {\bibfnamefont {C.}~\bibnamefont {Casciola}},\ }\bibfield  {title}
  {\enquote {\bibinfo {title} {Curvature effects in turbulent premixed flames
  of h2/air: a dns study with reduced chemistry},}\ }\href@noop {} {\bibfield
  {journal} {\bibinfo  {journal} {Flow, Turbulence and Combustion}\ }\textbf
  {\bibinfo {volume} {94}},\ \bibinfo {pages} {359--379} (\bibinfo {year}
  {2015})}\BibitemShut {NoStop}%
\bibitem [{\citenamefont {Li}\ and\ \citenamefont {Laizet}(2010)}]{2decomp}%
  \BibitemOpen
  \bibfield  {author} {\bibinfo {author} {\bibfnamefont {N.}~\bibnamefont
  {Li}}\ and\ \bibinfo {author} {\bibfnamefont {S.}~\bibnamefont {Laizet}},\
  }\bibfield  {title} {\enquote {\bibinfo {title} {2decomp \& fft-a highly
  scalable 2d decomposition library and fft interface},}\ }in\ \href@noop {}
  {\emph {\bibinfo {booktitle} {Cray User Group 2010 conference}}}\ (\bibinfo
  {year} {2010})\ pp.\ \bibinfo {pages} {1--13}\BibitemShut {NoStop}%
\bibitem [{\citenamefont {Pope}(2001)}]{pope2001turbulent}%
  \BibitemOpen
  \bibfield  {author} {\bibinfo {author} {\bibfnamefont {S.~B.}\ \bibnamefont
  {Pope}},\ }\href@noop {} {\enquote {\bibinfo {title} {Turbulent flows},}\ }
  (\bibinfo {year} {2001})\BibitemShut {NoStop}%
\bibitem [{\citenamefont {Luchini}(2017)}]{Luchini2017}%
  \BibitemOpen
  \bibfield  {author} {\bibinfo {author} {\bibfnamefont {P.}~\bibnamefont
  {Luchini}},\ }\bibfield  {title} {\enquote {\bibinfo {title} {Universality of
  the turbulent velocity profile},}\ }\href@noop {} {\bibfield  {journal}
  {\bibinfo  {journal} {Phys. Rev. Lett.}\ }\textbf {\bibinfo {volume} {118}},\
  \bibinfo {pages} {224501} (\bibinfo {year} {2017})}\BibitemShut {NoStop}%
\bibitem [{\citenamefont {Reynolds}\ and\ \citenamefont
  {Hussain}(1972)}]{reynolds1972mechanics}%
  \BibitemOpen
  \bibfield  {author} {\bibinfo {author} {\bibfnamefont {W.~C.}\ \bibnamefont
  {Reynolds}}\ and\ \bibinfo {author} {\bibfnamefont {A.}~\bibnamefont
  {Hussain}},\ }\bibfield  {title} {\enquote {\bibinfo {title} {The mechanics
  of an organized wave in turbulent shear flow. part 3. theoretical models and
  comparisons with experiments},}\ }\href@noop {} {\bibfield  {journal}
  {\bibinfo  {journal} {Journal of Fluid Mechanics}\ }\textbf {\bibinfo
  {volume} {54}},\ \bibinfo {pages} {263--288} (\bibinfo {year}
  {1972})}\BibitemShut {NoStop}%
\bibitem [{\citenamefont {Philip}(1972{\natexlab{a}})}]{philip1972integral}%
  \BibitemOpen
  \bibfield  {author} {\bibinfo {author} {\bibfnamefont {J.~R.}\ \bibnamefont
  {Philip}},\ }\bibfield  {title} {\enquote {\bibinfo {title} {Integral
  properties of flows satisfying mixed no-slip and no-shear conditions},}\
  }\href@noop {} {\bibfield  {journal} {\bibinfo  {journal} {Zeitschrift
  f{\"u}r angewandte Mathematik und Physik ZAMP}\ }\textbf {\bibinfo {volume}
  {23}},\ \bibinfo {pages} {960--968} (\bibinfo {year}
  {1972}{\natexlab{a}})}\BibitemShut {NoStop}%
\bibitem [{\citenamefont {Philip}(1972{\natexlab{b}})}]{philip1972flows}%
  \BibitemOpen
  \bibfield  {author} {\bibinfo {author} {\bibfnamefont {J.~R.}\ \bibnamefont
  {Philip}},\ }\bibfield  {title} {\enquote {\bibinfo {title} {Flows satisfying
  mixed no-slip and no-shear conditions},}\ }\href@noop {} {\bibfield
  {journal} {\bibinfo  {journal} {Zeitschrift f{\"u}r Angewandte Mathematik und
  Physik (ZAMP)}\ }\textbf {\bibinfo {volume} {23}},\ \bibinfo {pages}
  {353--372} (\bibinfo {year} {1972}{\natexlab{b}})}\BibitemShut {NoStop}%
\bibitem [{\citenamefont {Lauga}\ and\ \citenamefont {Stone}(2003)}]{lauga}%
  \BibitemOpen
  \bibfield  {author} {\bibinfo {author} {\bibfnamefont {E.}~\bibnamefont
  {Lauga}}\ and\ \bibinfo {author} {\bibfnamefont {H.~A.}\ \bibnamefont
  {Stone}},\ }\bibfield  {title} {\enquote {\bibinfo {title} {Effective slip in
  pressure-driven stokes flow},}\ }\href@noop {} {\bibfield  {journal}
  {\bibinfo  {journal} {Journal of Fluid Mechanics}\ }\textbf {\bibinfo
  {volume} {489}},\ \bibinfo {pages} {55--77} (\bibinfo {year}
  {2003})}\BibitemShut {NoStop}%
\bibitem [{\citenamefont {Fukagata}, \citenamefont {Iwamoto},\ and\
  \citenamefont {Kasagi}(2002)}]{fuka_bil}%
  \BibitemOpen
  \bibfield  {author} {\bibinfo {author} {\bibfnamefont {K.}~\bibnamefont
  {Fukagata}}, \bibinfo {author} {\bibfnamefont {K.}~\bibnamefont {Iwamoto}}, \
  and\ \bibinfo {author} {\bibfnamefont {N.}~\bibnamefont {Kasagi}},\
  }\bibfield  {title} {\enquote {\bibinfo {title} {Contribution of reynolds
  stress distribution to the skin friction in wall-bounded flows},}\
  }\href@noop {} {\bibfield  {journal} {\bibinfo  {journal} {Physics of
  Fluids}\ }\textbf {\bibinfo {volume} {14}},\ \bibinfo {pages} {L73--L76}
  (\bibinfo {year} {2002})}\BibitemShut {NoStop}%
\end{thebibliography}%

\end{document}